\documentclass[journal,comsoc]{IEEEtran}

\usepackage{cite}
\usepackage{graphicx}
\usepackage{CJK}
\usepackage{amsmath}
\usepackage{amssymb}
\usepackage{textcomp}
\usepackage{pifont}
\usepackage{dblfloatfix}
\usepackage{cancel}
\usepackage[linewidth=0.5pt,linecolor=blue]{mdframed}

\newtheorem{theorem}{Theorem}

\newtheorem{lemma}{Lemma}{}
{}
\newtheorem{remark}{Remark}{}
\newtheorem{definition}{Definition}{}
\begin{document}

\title{
Analysis of Intelligent Reflecting Surface-Enhanced Mobility Through a Line-of-Sight State Transition Model
}

\markboth{}%
{Shell \MakeLowercase{\textit{et al.}}: Equivalent Modelling and Analysis of Handover Process in  IRS-assisted networks }

\author{Haoyan~Wei,~\IEEEmembership{Member,~IEEE,} and Hongtao~Zhang,~\IEEEmembership{Senior~Member,~IEEE}
\thanks{The authors are with the Key Lab of Universal Wireless
Communications, Ministry of Education of China, Beijing University of Posts
and Telecommunications, Beijing 100876, China (e-mail: weihaoyan@bupt.edu.cn; htzhang@bupt.edu.cn).

This work has been submitted to the IEEE for possible publication. Copyright may be transferred without notice, after which this version may no longer be accessible.}
}

\maketitle

\begin{abstract}
Rapid signal fluctuations due to blockage effects cause excessive handovers (HOs) and degrade mobility performance. By reconfiguring line-of-sight (LoS) Links through passive reflections, intelligent reflecting surface (IRS) has the potential to address this issue. Due to the lack of introducing blocking effects, existing HO analyses cannot capture excessive HOs or exploit enhancements via IRSs. This paper proposes an LoS state transition model enabling analysis of mobility enhancement achieved by IRS-reconfigured LoS links, where LoS link blocking and reconfiguration utilizing IRS during user movement are explicitly modeled as stochastic processes. Specifically, the condition for blocking LoS links is characterized as a set of possible blockage locations, the distribution of available IRSs is thinned by the criteria for reconfiguring LoS links. In addition, neighboring BSs are categorized by probabilities of LoS states to enable HO decision analysis. By projecting distinct gains of LoS states onto a uniform equivalent distance criterion, mobility enhanced by IRS is quantified through the compact expression of HO probability. Results show the probability of dropping into non-LoS due to movement decreases by 70\% when deploying IRSs with the density of \boldmath{$\rm{ 93/km^2}$}, and HOs decrease by 57\% under the optimal IRS distributed deployment parameter.
\end{abstract}


\IEEEpeerreviewmaketitle

\vspace{-0.5cm}
\section{Introduction}

\IEEEPARstart {U}{tilization} of higher frequency bands in 5G networks and beyond allows for larger bandwidths and data rates; however, vulnerability of the wireless link to blockages is increased \cite{b1}. Frequent channel quality degradation caused by blockage effects leads to excessive handovers (HOs), which result in extra latency, signaling overhead, user equipment (UE) power consumption, and the risk of link failures \cite{b2}. With the capability to reconfigure line-of-sight (LoS) links, the intelligent reflecting surface (IRS) has emerged as a candidate technology for addressing this degradation in mobility performance.

For the blocked link, an IRS can provide an indirect LoS path between the base station (BS) and the UE allowing the signal to \emph{bypass} the blockage. Equipped with $N$ independent reflective units, the IRS can achieve $O(N^2)$ reflection gain \cite{b3,UAVIRS}. Therefore, IRS is expected to maintain a fine signal strength and avoid excessive HOs caused by the sharp signal drop from link blocking.
Based on the above discussion, the idea of enhancing mobility performance by exploiting IRS is inspired, and insightful design guidelines are desired.
However, the transition between LoS and non-LoS (NLoS) features randomness owing to the random distribution of blockages and the random movement of users. Additionally, the LoS link reconfiguration is affected by the LoS states of BS-IRS and IRS-UE links and the validity of the reflective path \cite{b4, b5}. Several analytical models for IRS-enhanced coverage under blockage effects have been proposed (e.g., \cite{b22, b23, b23.5}).  However, \cite{b22, b23, b23.5}  consider \emph{stationary} users and focus on transmission metrics, HO analysis of IRS-aided networks remains in its infancy. Since existing HO models lack modeling of LoS state transitions and IRS-reconfigured LoS links, the development of an analytical HO model for mining IRS-enhanced mobility under blockage effects remains an open issue.

\vspace{-0.3cm}
\subsection{Related Works}
\vspace{-0.1cm}
To evaluate the HO in networks and provide insightful design guidelines, stochastic geometry has been extensively employed, where HO probability, HO failure and ping-pong effects (e.g., \cite{b23.6}), and sojourn time (e.g., \cite{b14}) are usually concerned. Note that the focus of this paper is on HO probability analysis. In \cite{b7,b8,b9}, the HO in networks with a single-tier BS was analyzed, where the HO trigger locations were modeled as sets of points equidistant from BSs because the transmission power and channel fading were assumed to be the same. However, blockage effects lead to differential path loss of LoS/NLoS \cite{b10}, and IRS brings reflection gain. Thus a closer BS may not provide a stronger signal. Therefore, the distance-based HO models in \cite{b7,b8,b9} are not applicable to IRS-aided networks with blockage effects.

Because UE-BS Euclidean distance is insufficient to determine the HO locations in several scenarios, HO analytical models based on the received signal strength (RSS) have been proposed. Equivalent analysis techniques \cite{b11,b12} and analytical geometric frameworks \cite{b13} have also been used in heterogeneous networks. The HO locations in networks with different BS transmission powers were proven to be circular boundaries in \cite{b145}. Circular HO boundaries were also adopted in \cite{b15,Fading} to analyze HO probability. However, the serving link may be obstructed by blockages during user movement, particularly in hotspots such as urban areas, and abrupt signal degradation leads to excessive HOs \cite{b17}, which is expected to be addressed by reconfigured LoS links provided by the IRS.
The HO models above consider only the negative exponential path loss between UE and BS without translations between LoS/NLoS links and IRS reflections; hence, the methods in \cite{b11,b12,b13} and circular boundary models \cite{b145,b15, Fading} are invalid.

To study LoS/NLoS conditions of a link with user movement, the intervals of LoS and NLoS links on the user trajectory were obtained in \cite{b18} without LoS/NLoS translation and HO analysis. HO analyses in scenarios with blockages were performed based on multi-directional \cite{b19}, dual-slope \cite{b16}, and average-weighted \cite{b20} path-loss models. However, statistical channel models cannot capture abrupt signal degradation, and their parameters are based on simulation fitting or artificial configurations that cannot explicitly introduce blockage parameters. Blockages were modeled exactly in \cite{b21} for the HO probability analysis. However, only thinned LoS BS density was considered in \cite{b21}, assuming that the LoS/NLoS transition analysis was not tractable. Moreover, the utilization of IRS was not considered in \cite{b16, b18,b19,b20,b21}. 
To mine the IRS gains on mobility performance, the reduction in HOs was formulated as an \emph{optimization problem} in \cite{b24} where a specific algorithm was presented instead of an analytical model. HO models considering IRSs were proposed to analyze probabilities of HO \cite{b25}, HO failure and ping-pong \cite{b25.5}. However, blockage effects were ignored in \cite{b25} and \cite{b25.5}.

In this context, regardless of HO studies with LoS/NLoS links \cite{b16,b19,b20,b21} or IRS channels \cite{b24,b25,b25.5}, there are still many unexplored issues in establishing an analytical model for HO enhancement by exploiting IRS-reconfigured LoS links.

\begin{itemize}
\item LoS state transitions with user movement have not been analyzed theoretically considering blockage modeling. LoS/NLoS path loss models are adopted in \cite{b19,b16,b20}. Nevertheless, statistical models cannot capture the signal degradation from abrupt link blocking, which is the main cause of excessive HOs. Although \cite{b21} conducts blockage modeling, the analysis of LoS/NLoS transitions is still omitted. Additionally, the new paradigm of IRS-reconfigured LoS links is not considered in \cite{b19,b16,b20,b21}.

\item Reconfigured LoS links via IRS have not been modeled theoretically in HO analysis. IRS channels are introduced into the HO analysis in \cite{b25} and  \cite{b25.5}. Nevertheless, all links in \cite{b25} and  \cite{b25.5} are assumed as LoS. HO models in \cite{b16,b19,b20,b21} only consider LoS/NLoS states of BS-user links. However, the establishment of reconfigured LoS links involves analyzing the LoS states of the BS-IRS and IRS-user link and the validity of the reflective path.

\item  Existing works have not obtained reference-worthy results to study the effect of IRS-reconfigured LoS links on HO performance.
Although results of HO performance without LoS/NLoS links were obtained in \cite{b25} and  \cite{b25.5}, the gain of LoS link reconfigurations via IRSs remains to be explored. Simulation-based results were realized in \cite{b24} for IRS-aided networks with blockages. However, simulations are computationally intensive and specific to a certain setting, making it difficult to obtain guidelines.

\end{itemize}

\begin{figure}[t]
\vspace{-0.2cm}
\centerline{\includegraphics[width=0.8\linewidth]{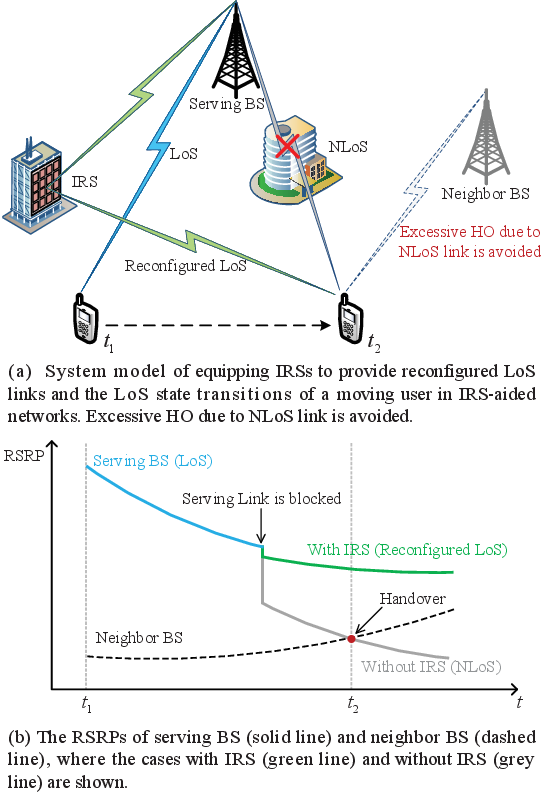}}
\vspace{-0.2cm}
\caption{System model for handovers in IRS-aided networks: (a) IRS network structure; (b) Changes in received signal strength with time.}
\vspace{-0.7cm}
\label{fig1}
\end{figure}

\vspace{-0.4cm}
\subsection{Contribution}
\vspace{-0.1cm}

This paper proposes an analytical framework for probabilities of the LoS state transition and HO in IRS-aided networks with blockages to explore the mobility enhancement achieved by IRS-reconfigured LoS links. The impact of IRS configurations on network mobility performance are presented along with insights and design guidelines.
The main contributions are summarized as follows.

\begin{itemize}
\item LoS state transitions between LoS, NLoS, and reconfigured LoS are theoretically analyzed with blockage modeling. To analyze the transition and reconfiguration of LoS links, the transition condition of blocking LoS links is characterized as a set of possible blockage locations, and available IRSs are determined by the criteria for reconfiguring LoS links. Probabilities of LoS state transitions with user movement in IRS-aided networks are derived.

\item The IRS-reconfigured LoS link is introduced into the HO analysis. The location distribution of the IRS that can reconfigure the LoS link for the user is generated by thinning all IRSs according to the IRS availability probability. With the aid of thinned IRS density, the IRS reflection gain is obtained based on the distance distribution between the user and its serving IRS, which is introduced into the HO decision analysis.

\item To explore enhancement of LoS link reconfigurations, the HO probability of IRS-aided networks is analyzed. The neighboring BSs are categorized depending on the probabilities of LoS states. The distinct gains of BSs with different LoS states are projected to a uniform equivalent distance criterion. Therefore, mobility performance enhanced by IRS is quantified by deducing the compact expression of HO probability.

\item Design insights obtained from the main results include: (i) the probability of blocking the serving link due to user movement is reduced by 70\% when setting the IRS density to $\rm{ 93/km^2}$ and the serving distance to 100m, which corresponds to the gain in BS density from $\rm{ 10/km^2}$ to $\rm{ 100/km^2}$; (ii) there is an optimal distributed deployment parameter that minimizes the probability of HO as the blocking effect is not severe, and conversely a more distributed IRS deployment option achieves a lower probability of HO.
\end{itemize}

\begin{figure*}[hb]
\vspace{-0.2cm}
\hrulefill
\begin{equation} \label{PIL}
\setlength\abovedisplayskip{4pt}
\setlength\belowdisplayskip{4pt}
\begin{aligned} 
&{p_i}\left( {r,\theta \left| d \right.} \right) = \frac{{{e^{ - \frac{{2{\lambda _o}}}{\pi }{\mathbb E}\left[ l \right]\left( {{d}' + {r}} \right)}}}}{{1 - {e^{ - \frac{{2{\lambda _o}}}{\pi }{\mathbb E}\left[ l \right]{d}}}}}\left( {{e^{\frac{{{\lambda _o}}}{{2\pi }}{\mathbb E}\left[ {{l^2}} \right]\left( {\frac{1}{2} + \frac{{\pi  - {\theta _1}}}{2}\cot {\theta _1}} \right)}} - {e^{ - \frac{{2{\lambda _o}}}{\pi }{\mathbb E}\left[ l \right]{d} + \frac{{{\lambda _o}}}{{2\pi }}{\mathbb E}\left[ {{l^2}} \right]\left[ {\sum\limits_{c = 1}^3 {\left( {\frac{1}{2} + \frac{{\pi  - {\theta _c}}}{2}\cot {\theta _c}} \right)} } \right]}}} \right)\left( {1 - \frac{{{\theta _1}}}{\pi }} \right),\\
&{\theta _1} = \arccos \left( {\frac{{r - d\cos \theta }}{{\sqrt {{d^2} + {r^2} - 2dr\cos \theta } }}} \right), \ \ \  {\theta _2} = \arccos \left( {\frac{{d - r\cos \theta }}{{\sqrt {{d^2} + {r^2} - 2dr\cos \theta } }}} \right),\ \ \  {\theta _3} = \theta , \ \ \ d' = \sqrt {{d^2} + {r^2} - 2dr\cos \theta } . 
\end{aligned}\tag{5}
\end{equation}
\begin{equation} \label{PIL2}
\setlength\abovedisplayskip{4pt}
\setlength\belowdisplayskip{4pt}
\begin{small}
\begin{aligned} 
&{\widehat p_i}\!\left( {{r_s},\!{r_e},\!d} \right) \!=\! \frac{1}{2}\!\left( {\left( {\frac{{\pi \max \left\{ {{r_s},{\mathbb E}\left[ l \right]} \right\}}}{{2{\lambda _o}{\mathbb E}\left[ l \right]}} \!+\! {{\left( {\frac{\pi }{{2{\lambda _o}{\mathbb E}\left[ l \right]}}} \right)}^2}} \right)\!{e^{ - \frac{{2{\lambda _o}{\mathbb E}\left[ l \right]}}{\pi }\left( {\max \left\{ {{r_s},{\mathbb E}\left[ l \right]} \right\} + d} \right)}} \!\!-\! \left( {\frac{{\pi \max \left\{ {{r_e},{\mathbb E}\left[ l \right]} \right\}}}{{2{\lambda _o}{\mathbb E}\left[ l \right]}} \!+\! {{\left( {\frac{\pi }{{2{\lambda _o}{\mathbb E}\left[ l \right]}}} \right)}^2}} \right)\!{e^{ - \frac{{2{\lambda _o}{\mathbb E}\left[ l \right]}}{\pi }\left( {\max \left\{ {{r_e},{\mathbb E}\left[ l \right]} \right\} + d} \right)}}} \right).
\end{aligned}\tag{6}
\end{small}
\end{equation}
\vspace{-0.7cm}
\end{figure*}

\vspace{-0.2cm}
\section{System Model}
\vspace{-0.1cm}
We consider a large-scale cellular network constituting BSs, users, and IRSs with blocking factored in. Users move in the network and hand over BSs based on their received signal strengths. The IRS reconfigures the LoS link by passive reflection from the BS blocked to avoid frequent HOs.

\vspace{-0.2cm}
\subsection{Network Model}
\vspace{-0.2cm}

In the IRS-aided wireless network considered, BSs are distributed according to a 2-dimensional homogenous Poisson point process (HPPP) with density $\lambda_b$ and the same transmission power $p_t$, denoted as $\Phi_b$. In terms of tractability and good fitness to the actual environment, the line Boolean model is adopted to model the distribution of blockages \cite{b6, b19, b22, b23}. In particular, the blockages are modeled as line segments of length $l$ and angle $\beta$. The locations of the midpoints of the blockages are modeled as an HPPP $\Phi_o$ with density ${\lambda _o}$. For any blockage denoted by ${o_k} \in {\Phi _o}$, the variable $l$ follows a uniform distribution within the range of $l_{\min}$ to $l_{\max}$. The variable $\beta$
is uniformly distributed between 0 and $2\pi$, which is the angle between the blockage and the positive direction of the $x$-axis.

As in \cite{b22,b23}, for the IRS distribution, a subset ${\Phi _i} \subset {\Phi _o}$ of blockages is equipped with IRSs on both sides and the density of ${\Phi _i}$ is $\lambda_i \!=\! \mu \lambda_o$ $\left(  0 \le \mu  \le 1  \right)$, where the value of $\mu$ represents the percentage of the blockages equipped with IRSs, and each IRS is equipped with $N$-tunable reflecting elements to assist in BS-user communications.\footnote{Like most models of IRS-aided networks \cite{b22,b23,b3}, we consider the IRS uses all $N$ elements to serve the typical user. Other cases can be easily analyzed by scaling $N$ (e.g., the IRS splits elements to serve multiple users) or $\lambda_i$ (e.g., IRSs have a probability of being occupied by other users).}

\vspace{-0.4cm}
\subsection{Channel Model}
\vspace{-0.1cm}
As in  \cite{b23.7}, Nakagami-$m$ fading is considered in this paper.\footnote{Nakagami-$m$ is a general distribution to model different channel fadings via the shape parameter $m$, e.g., it is equivalent to the Rayleigh fading if $m\!=\!1$ and is closely approximated as the Rician fading with parameter $K$ if $m \!=\! {\left( {K \!+\! 1} \right)^2}/\left( {2K \!+\! 1} \right)$ \cite{b23.8}.}
Because the filtering process is performed in the channel measurement at the user terminal, the effect of the fast fading of the channel is averaged \cite{b25, b14, b203}.
Therefore, for LoS and NLoS links, the measured signal strength from BS $m$ is given by
\begin{equation}
\setlength\abovedisplayskip{0pt}
\setlength\belowdisplayskip{1pt}
  {P_k}\left( {{d_0}} \right) = {\mathbb E} \left[ {{p_t}{h_k}{K_k}d_0^{ - {\alpha _k}}} \right] \mathop  = \limits^{\left( a \right)}  {p_t}{K_k}d_0^{ - {\alpha _k}},k \in \left\{ {L,N} \right\},
\end{equation}
where $K_L$ and $K_N$ represent the additional path losses in LoS and NLoS links, $d_0$ is the distance between the user and the BS $m$, ${\alpha _{L}}$ and ${\alpha _{N}}$ are the path loss exponents in LoS and NLoS links, $h_{L}$ and $h_{N}$ are fading coefficients in LoS and NLoS links with parameters $m_L$ and $m_N$, (a) follows ${\mathbb E}\left[ {{h_k}} \right] = 1$.

As in \cite{b25, b3, b23.7}, it is assumed that IRS adjusts the phase shift such that the $N$ reflected signals are combined at the aligned phase at its served UE for maximum signal strength. Therefore, for the reconfigured LoS links, the measured signal strength from BS $m$ is given by
\begin{equation}
\setlength\abovedisplayskip{2pt}
\setlength\belowdisplayskip{2pt}
\begin{small}
\begin{aligned}
& {P_R}\left( {{d'_0},{r_0}} \right) = {\mathbb E}\left[ {{p_t}K_L^2 r_0^{ - {\alpha _L}}{d'_0}^{ - {\alpha _L}}{{\left( {\sum\limits_{n = 1}^N {\sqrt {h_{L,n}^{b - r}} \sqrt {h_{L,n}^{r - u}} } } \right)}^2}} \right] \\
 &  \ \ \ \ \ \ \ \ \ \ \ \ \ \ = {p_t}K_L^2{G_{bf}}r_0^{ - {\alpha _L}}{d'_0}^{ - {\alpha _L}},
\end{aligned}
\end{small}
\end{equation}
where
\begin{equation}
\begin{small}
\setlength\abovedisplayskip{2pt}
\setlength\belowdisplayskip{0pt}
\begin{aligned} 
{G_{\! bf}} \!=\! {\mathbb E}\!\left[ {{{\left( {\sum\limits_{n = 1}^N \!{\sqrt {h_{L,n}^{b - r}}\! \sqrt {h_{L,n}^{r - u}} } } \right)}^{\! 2}}} \right] \!\!=\! {N^2} \!+\! N\left( \!{1 \!-\! \frac{1}{{m_L^2}}{{\left( {\frac{{\Gamma\! \left( {{m_L} \!+\! \frac{1}{2}} \right)}}{{\Gamma\! \left( {{m_L}} \right)}}} \right)}^4}} \right)
\end{aligned}
\end{small}
\end{equation}
is the reflecting gain under Nakagami-$m$ fading $\rm{[32, Eq.(22b)]}$, $\Gamma \! \left(  \cdot  \right)$ is the gamma function, ${h_{L,n}^{b - r}}$ and ${h_{L,n}^{r - u}}$ are fading coefficients of channels from BS to $n$-th IRS element and from $n$-th IRS element to user with the same parameter $m_L$, $r_0$ is the IRS--user distance, ${d’_0}$  is the IRS--BS distance, and the NLoS component is omitted because it is far lower than the component of the IRS reflection \cite{b23}. 

It is worth noting that in the HO probability analysis, interference from neighboring BSs is not considered since the reference signals from BSs used for HO decisions are orthogonal \cite{b23.9}.

\begin{remark}
The analytical framework proposed can be easily extended to scenarios considering the hardware impairments of IRS by modifying Eq. (2) as need, e.g., refer to \cite{R1} and \cite{R2}. To focus on the topic, the hardware impairments are not discussed further, which gives an upper bound on the performance.
\end{remark} %\cite{R1} and \cite{R2}

\begin{figure}[t]
\centerline{\includegraphics[width=0.85\linewidth]{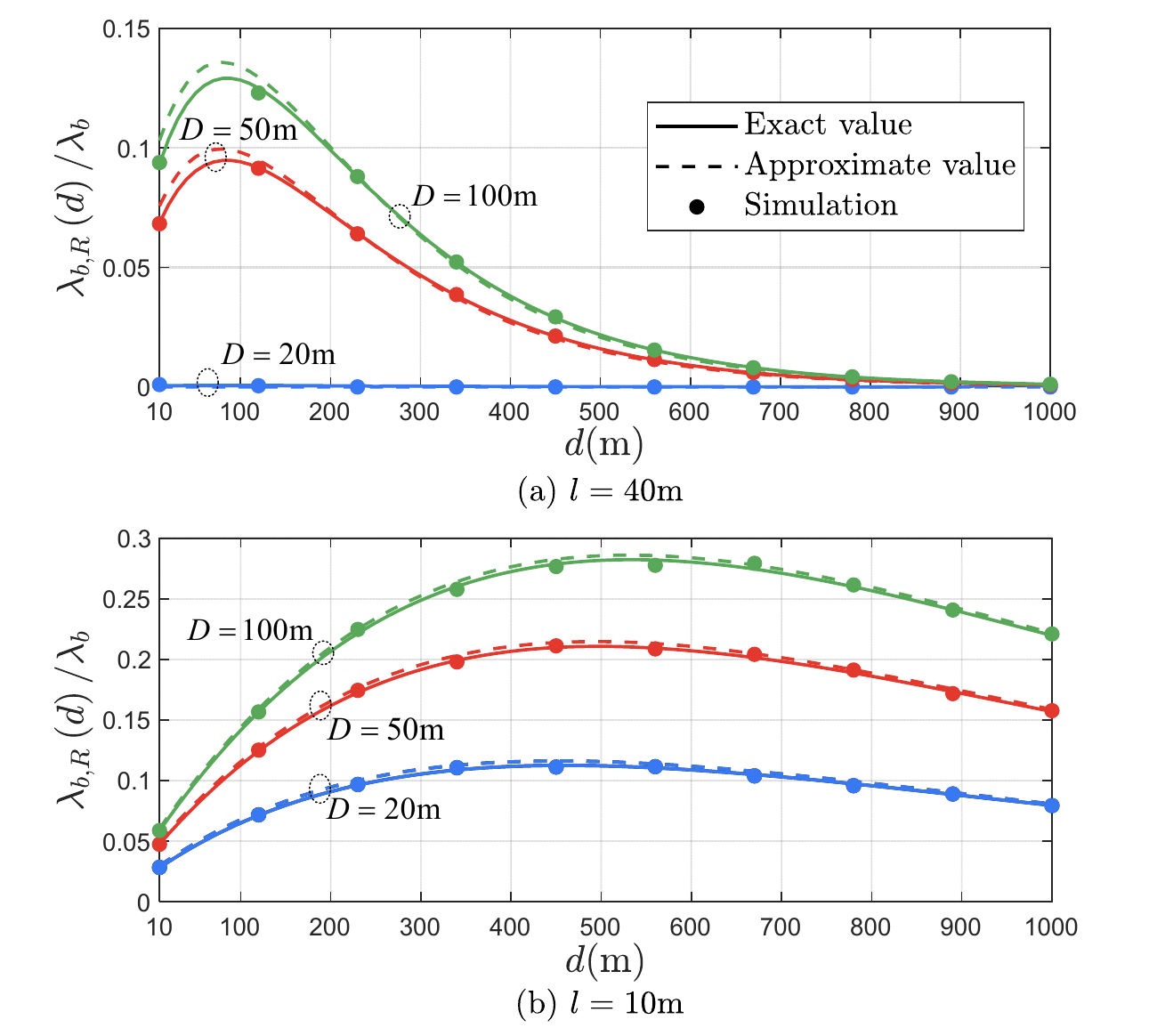}}
\vspace{-0.3cm}
\caption{Validation of the BS density in \eqref{lambda} considering different UE-BS distance $d$, IRS serving distance $D$, and length of blockage $l$. Other related parameters are given by, $\lambda_b =10\rm{/km^2}$, $\lambda_o =500\rm{/km^2}$, and $\mu = 0.5$.}
\vspace{-0.5cm}
\label{fig1}
\end{figure}

\begin{figure*}[b]
\vspace{-0.2cm}
\hrulefill
\begin{equation} \label{Ak}
\setlength\abovedisplayskip{4pt}
\setlength\belowdisplayskip{4pt}
\begin{aligned} 
&{{\cal A}_L} = {{\mathbb E}_{{d^L}}}\left[ {\exp \left\{ { - 2\pi \left[ {\int_0^{{{\widetilde d}^N}} \!\!\! {{\lambda _{b,N}}\left( {{d^N}} \right){d^N}{\rm{d}}{d^N}}  + \int_0^\infty \!\!\! {{\lambda _{b,R}}\left( {{d^R}} \right){{\cal F}_{{r_1}}}\left( {{{\widetilde r}^L}\left| {{d^R}} \right.} \right){d^R}{\rm{d}}{d^R}} } \right]} \right\}} \right],\\
&{{\cal A}_N} = {{\mathbb E}_{{d^N}}}\left[ {\exp \left\{ { - 2\pi \left[ {\int_0^{{{\widetilde d}^L}} \!\!\!{{\lambda _{b,L}}\left( {{d^L}} \right){d^L}{\rm{d}}{d^L}}  + \int_0^\infty \!\!\! {{\lambda _{b,R}}\left( {{d^R}} \right){{\cal F}_{{r_1}}}\left( {{{\widetilde r}^N}\left| {{d^R}} \right.} \right){d^R}{\rm{d}}{d^R}} } \right]} \right\}} \right],\\
&{{\cal A}_R} = {{\mathbb E}_{{d^R}}}\left[ {\exp \left\{ { - 2\pi \left[ {\int_0^\infty \!\!\! {{\lambda _{b,L}}\left( {{d^L}} \right)\left[ {1 - {{\cal F}_{{r_1}}}\left( {{{\widetilde r}^L}\left| {{d^R}} \right.} \right)} \right]{d^L}{\rm{d}}{d^L}}  + \int_0^\infty \!\!\! {{\lambda _{b,N}}\left( {{d^N}} \right)\left[ {1 - {{\cal F}_{{r_1}}}\left( {{{\widetilde r}^N}\left| {{d^R}} \right.} \right)} \right]{d^N}{\rm{d}}{d^N}} } \right]} \right\}} \right],\\
&{{\widetilde d}^N} \!=\! {\left( {\frac{{{K_N}}}{{{K_L}}}} \right)^{\frac{1}{{{\alpha _N}}}}}{\left( {{d^L}} \right)^{\frac{{{\alpha _L}}}{{{\alpha _N}}}}},\ \ {{\widetilde d}^L}  \!=\! {\left( {\frac{{{K_L}}}{{{K_N}}}} \right)^{\frac{1}{{{\alpha _L}}}}}{\left( {{d^N}} \right)^{\frac{{{\alpha _N}}}{{{\alpha _L}}}}},\ \ {{\widetilde r}^L}  \!=\! {\left( {{K_L}{G_{bf}}} \right)^{\frac{1}{{{\alpha _L}}}}}\frac{{{d^L}}}{{{d^R}}},\ \ {{\widetilde r}^N}  \!=\! {\left( {\frac{{K_L^2{G_{bf}}}}{{{K_N}}}} \right)^{\frac{1}{{{\alpha _L}}}}}\frac{{{{\left( {{d^N}} \right)}^{\frac{{{\alpha _N}}}{{{\alpha _L}}}}}}}{{{d^R}}}, 
\\&{f_{{d^k}}}\left( {{d^k}} \right) = 2\pi {\lambda _{b,k}}\left( {{d^k}} \right){d^k}\exp \left\{ { - 2\pi \int_0^{{d^k}} {{\lambda _{b,k}}\left( d \right)d{\rm{d}}d} } \right\},k \in \left\{ {L,N,R} \right\}.
\end{aligned}\tag{10}
\end{equation}

\begin{equation} \label{Fx0}
\setlength\abovedisplayskip{4pt}
\setlength\belowdisplayskip{4pt}
\begin{aligned} 
{\Xi _k}\left( {{d^k}} \right) = \left\{ \begin{array}{ll}
\exp \left\{ { - 2\pi \left( {\int\limits_0^{{d^L}} {{\lambda _{b,L}}\left( {d} \right){d}{\rm{d}}{d}}  + \int\limits_0^{{{\widetilde d}^N}} {{\lambda _{b,N}}\left( {d} \right){d}{\rm{d}}{d}}  + \int\limits_0^\infty  {{\lambda _{b,R}}\left( {d^R} \right){{\cal F}_{{r_1}}}\left( {{{\widetilde r}^L}\left| {d^R} \right.} \right){d^R}{\rm{d}}{d^R}} } \right)} \right\},&\!\!\!k = L\\
\exp \left\{ { - 2\pi \left( {\int\limits_0^{{d^N}} {{\lambda _{b,N}}\left( d \right)d{\rm{d}}d}  + \int\limits_0^{{{\widetilde d}^L}} {{\lambda _{b,L}}\left( d \right)d{\rm{d}}d}  + \int\limits_0^\infty  {{\lambda _{b,R}}\left( d^R \right){{\cal F}_{{r_1}}}\left( {{{\widetilde r}^N}\left| d^R \right.} \right)d^R{\rm{d}}d^R} } \right)} \right\},&\!\!\!k = N\\
\exp \left\{ { - 2\pi \left( {\int\limits_0^{{d^R}}\!\!{{\lambda _{b,R}}\left( d \right)d{\rm{d}}d}  \!+\! \int\limits_0^\infty  \!\!{{\lambda _{b,L}}\left( {{d^L}} \right)\!\!\left( {1 \!-\! {{\cal F}_{{r_1}}}\!\!\left( {{\widetilde r^L}\left| {{d^R}} \right.} \right)} \right){d^L}{\rm{d}}{d^L}}  \!+\!\! \int\limits_0^\infty \!\!{{\lambda _{b,N}}\!\left( {{d^N}} \right)\!\left( {1 \!-\! {{\cal F}_{{r_1}}}\left( {{\widetilde r^N}\left| {{d^R}} \right.} \right)} \right){d^N}{\rm{d}}{d^N}} } \right)} \right\},&\!\!\!k = R
\end{array} \right.
\end{aligned}\tag{12}
\end{equation}

\vspace{-0.3cm}
\end{figure*}

\vspace{-0.4cm}
\subsection{Connection Policy}
\vspace{-0.15cm}
As in \cite{b22, b23}, we consider that the IRS is scheduled to provide reconfigured LoS link if the BS-user direct link is blocked. Moreover, a limited IRS serving distance $D$ is considered  \cite{b3}, i.e., the IRS only serves users within distance $D$ and visible. Therefore we summarize three possible LoS states of links in IRS-aided networks.
 
\begin{definition}
A LoS link exists when there are no blockages obstructing the path between the user and BS.
\end{definition}

\begin{definition}
A reconfigured LoS link exists when the path between the user and the BS is blocked but the user is within the serving distance of the IRS that can reflect the signal, and the paths between the user and the IRS and between the IRS and the BS are not blocked.
\end{definition}

\begin{definition}
An NLoS link exists when blockages obstruct the path between the user and BS and no IRS can build reconfigured LoS links for the user.
\end{definition}

The user associates with the BS that provides the maximum received signal power and performs intercell HOs via channel measurements.

\begin{remark}
As proved in \cite{R3}, the IRS serving distance $D$ implies that the IRS serves the user if the relative signal strength gain reaches a threshold, thus $D$  can be determined by pragmatically setting the threshold. Other IRS access strategies can be applied to the proposed framework by modifying $D$ as infinity, functions, random variables, etc.
\end{remark} %\cite{R3}

\vspace{-0.3cm}
\subsection{User Mobility Model}

Since it is obviously unreasonable for users to move through blockages, we modify the traditional random walk model \cite{R4} to account for the effects of blockages on user mobility. Specifically, in each unit of time, the typical user randomly selects a direction and is expected to move in that direction for a unit of time at a constant velocity $v$. However, if the selected direction includes a blockage to the trajectory of the user, the user reselects a direction until it will not be blocked. After the direction is determined, the user performs mobility, after which the above process is repeated. The effect of the modified model on the theoretical analysis will be illustrated in the corresponding derivations.  %\cite{R4}

\vspace{-0.2cm}
\section{Mobility Analysis}

To exploit the enhancement of IRS in the HO performance, we derive theoretical expressions for probabilities of LoS state transitions and HO in IRS-aided networks, as presented in this section. First, we obtain the distributions of BSs with different LoS states and IRSs that can reconfigure LoS links. Then, we analyze the LoS state of the link between the typical user and the BS connected at the start of the typical unit time. Afterward, the LoS state transitions of the serving link after the user moves a unit time are analyzed. Finally, the probability that the user triggers an HO is derived.

\vspace{-0.3cm}
\subsection{Distribution of BSs and IRSs}

Different LoS states exist between a typical user and BSs in the IRS-aided network. Without loss of generality, we consider a typical user to be located at the origin, the BSs are thinned into three types of BSs based on the LoS states with the typical user, denoted as ${\Phi _{b,L}}$, ${\Phi _{b,N}}$ and ${\Phi _{b,R}}$, corresponding to LoS, NLoS, and reconfigured LoS states. According to the definitions of LoS states in Section II.B, ${\Phi _{b,L}}$, ${\Phi _{b,N}}$ and ${\Phi _{b,R}}$ are mutually exclusive and ${\Phi _{b,L}} \cup {\Phi _{b,N}} \cup {\Phi _{b,R}} = {\Phi _b}$ \cite{b22, b23}. Then, we have the following lemma.

\begin{lemma}
For a typical user, BSs with LoS states of LoS, NLoS and reconfigured LoS follow HPPPs with densities of ${\lambda _{b,L}}\left( d \right)$, ${\lambda _{b,N}}\left( d \right)$, and ${\lambda _{b,R}}\left( d \right)$, respectively, where $d$ is the distance to the typical user. The expressions for the densities are given by
\begin{equation}
\setlength\abovedisplayskip{2pt}
\setlength\belowdisplayskip{2pt}
\begin{aligned} \label{lambda}
{\lambda _{b,L}}\!\left( d \right) \!&=\! {\lambda _b}{e^{ - {\lambda _o}\frac{2}{\pi }{\mathbb E}\left[ l \right]d}},\\[-2mm]
{\lambda _{b,N}}\!\left( d \right) \!&=\! {\lambda _b}\left( {1 - {e^{ - {\lambda _o}\frac{2}{\pi }{\mathbb E}\left[ l \right]d}}} \right){e^{ - {\lambda _i}\int\limits_{ - \pi }^\pi  {\int\limits_0^D {{p_i}\left( {r,\theta \left| d \right.} \right)r{\rm{d}}r{\rm{d}}\theta } } }}\\[-1mm]
&\  \approx {\lambda _b}\left( {1 - {e^{ - {\lambda _o}\frac{2}{\pi }{\mathbb E}\left[ l \right]d}}} \right){e^{ - {\lambda _i}{\widehat {{p}}_i}\left( 0,{D}, {d} \right)}},\\[-2mm]
{\lambda _{b,R}}\!\left( d \right) &\!=\! {\lambda _b}\left( {1 \!-\! {e^{ - {\lambda _o}\frac{2}{\pi }{\mathbb E}\left[ l \right]d}}} \right)\left( \!{1 \!-\! {e^{ - {\lambda _i}\!\!\int\limits_{ - \pi }^\pi \! {\int\limits_0^D {\!{p_i}\left( {r,\theta \left| d \right.} \right)r{\rm{d}}r{\rm{d}}\theta } } }}} \right)\\[-1mm]
&\  \approx {\lambda _b}\left( {1 - {e^{ - {\lambda _o}\frac{2}{\pi }{\mathbb E}\left[ l \right]d}}} \right)\left( {1 - {e^{ - {\lambda _i}{{\widehat p}_i}\left( 0,{D}, {d} \right)}}} \right),
\end{aligned}
\end{equation}
where ${p_i}\left( {r,\theta \left| d \right.} \right)$ is the probability that the IRS at a distance of $r$ from the user and an angle of $\theta$ relative to the user-BS link can build the reconfigured LoS link under a given $d$, which is given in \eqref{PIL} (on bottom of this page), and ${\widehat {{p}}_i}\!\left({r_s}, {r_e}, {d} \right)$ is the approximation of $\int_{ -\! \pi }^\pi \! {\int_{r_s}^{r_e} \!\!\!{{p_i}\!\left( {r,\theta \left| d \right.} \right)\!r{\rm{d}}r{\rm{d}}\theta } } $, which is given in \eqref{PIL2} (on bottom of this page).
\end{lemma}

\begin{IEEEproof}
See Appendix A.
\end{IEEEproof}

Fig. 2 validates the accuracy of the expressions for BS densities (obtained in Eq. \eqref{lambda}) with the Monte-Carlo simulations. The derived closed-form expressions also match the exact expressions well.

\begin{remark} 
From {\it Lemma 1}, it can be observed that the probability of being able to build the reconfigured LoS link, which is given by $ {1 \!-\! {e^{ - {\lambda _i}{{\widehat p}_i}\left( 0,{D}, {d} \right)}}} $, (i) decreases with $d$ with a diminishing rate and asymptotically approaches zero, (ii) increases with $\lambda_i$ with an exponentially decreasing rate, (iii) increases with $D$ with a decreasing rate and asymptotically approaches a finite limit, (iv) decreases with ${{\mathbb E}\left[ l \right]}$ and $\lambda_o$ exhibiting an exponentially decreasing trend and asymptotically approaching 0. In practical design, parameters must be simultaneously optimized (lowering $d$ and increasing $D$ and $\lambda_i$) to mitigate severe blockage effects (characterized by high ${{\mathbb E}\left[ l \right]}$ and $\lambda_o$), as the benefits of single-parameter improvements progressively diminish.
\end{remark}

\setcounter{equation}{6}

In the case of LoS state in the reconfigured LoS, we focus on the distribution of the distance between a typical user and its serving IRS. The network schedules an IRS that satisfies the serving condition and is closest to the user because it can provide the strongest received signal and maximizes the assurance that the IRS-user link remains unblocked afterwards.
\begin{lemma}
If the link between the user and the BS at a distance of $d$ is the reconfigured LoS, the cumulative distribution function (cdf) and probability density function (pdf) of the distance, ${r_1} \!\!\in\!\! \left[ {0,D} \right)$, between the user and its serving IRS are given by
\begin{equation} \label{Fr0}
\setlength\abovedisplayskip{3pt}
\setlength\belowdisplayskip{1pt}
\begin{small}
\begin{aligned}
{{\cal F}_{{r_1}}}\!\!\left( {{r_1}\left| d \right.} \right) &\!=\!  \frac{{1 \!-\! {e^{ - {\lambda _i}\int_{ - \pi }^\pi  {\int_0^{{r_1}}\!\! {{p_i}\left( {r,\theta \left| d \right.} \right)r{\rm{d}}r{\rm{d}}\theta } } }}}}{{1 \!-\! {e^{ - {\lambda _i}\int_{ - \pi }^\pi  {\int_0^D \!\!{{p_i}\left( {r,\theta \left| d \right.} \right)r{\rm{d}}r{\rm{d}}\theta } } }}}}  \approx\! \frac{{1 - {e^{ - {\lambda _i}{{\widehat p}_i}\left( {0,{r_1},d} \right)}}}}{{1 - {e^{ - {\lambda _i}{{\widehat p}_i}\left( {0,D,d} \right)}}}}, 
\end{aligned}
\end{small}
\end{equation}
\begin{equation}
\setlength\abovedisplayskip{0pt}
\setlength\belowdisplayskip{2pt}
\begin{small}
\begin{aligned}
{f_{{r_1}}}\!\!\left( {{r_1}\left| d \right.} \right) \!&=\! \frac{{{\lambda _i}{e^{ - {\lambda _i}\!\int_{ - \pi }^\pi  {\int_0^{{r_1}} {{p_i}\left( {r,\theta \left| d \right.} \right)r{\rm{d}}r{\rm{d}}\theta } } }}}}{{1 - {e^{ - {\lambda _i}\!  \int_{ - \pi }^\pi  {\int_0^D {{p_i}\left( {r,\theta \left| d \right.} \right)r{\rm{d}}r{\rm{d}}\theta } } }}}}\int_{ - \pi }^\pi \!\!\! {{p_i}\!\left( {{r_1},\!\theta \left| d \right.} \right)\!{r_1}{\rm{d}}\theta }  \\
& \approx \frac{{{\lambda _i}{e^{ - {\lambda _i}{{\widehat p}_i}\left( {0,{r_1},d} \right)}}}}{{1 - {e^{ - {\lambda _i}{{\widehat p}_i}\left( {0,D,d} \right)}}}}{\widehat p_i}^{\ \! \prime} \left( {0,{r_1},d} \right),\\[-3mm]
\end{aligned}
\end{small}
\end{equation}
where
\begin{equation}
\setlength\abovedisplayskip{0pt}
\setlength\belowdisplayskip{1pt}
\begin{small}
\begin{aligned}
&{\widehat p_i}^{\ \! \prime} \!\left( {0,{r_0},d} \right) \!=\! \frac{{\partial {{\widehat p}_i}\left( {0,{r_0},d} \right)}}{{\partial {r_0}}}  \\\! & \ \ \ \ \ \ \ \ \ \ \ \  =\! \left\{ \begin{array}{l}
 \!\!\!\! \left( {\frac{1}{2} \!+\! \frac{\pi }{{4{\lambda _o}{\mathbb E}\left[ l \right]}}} \right)\!{e^{ - \frac{{2{\lambda _o}{\mathbb E}\left[ l \right]}}{\pi }\left( {{r_0} + d} \right)}},{\mathbb E}\left[ l \right] < {r_0} \le D\\
 \!\!0,\ \ \ \ \ \ \ \ \ \ \ \ \ \ \ \ \ \ \ \ \ \ \ \ \ \ \ \ \ \ {\rm{others}}
\end{array} \right.
\end{aligned}\!\!\!.
\end{small}
\end{equation}
\end{lemma}

\begin{IEEEproof}
See Appendix B.
\end{IEEEproof}

\begin{remark}
From {\it Lemma 2}, the increase in $\lambda_i$ raises the probability of taking a smaller distance $r_1$, the increase in $D$ brings the extra probability of $r_1$ taking on a larger value, decreases in $\lambda_o$ and ${{\mathbb E}\left[ l \right]}$ cause $f_{r_1} \left( {{r_1}\left| d \right.} \right)$ to increase at both small and large $r_1$ values. Hence, while sufficient $D$ or less blocking ensures the probability of building reconfigured LoS links, significant gains are not guaranteed due to large $r_1$, which need to be resolved by lifting $\lambda_i$.
\end{remark}

\vspace{-0.3cm}
\subsection{Initial LoS State of Serving Link}

HO analysis focuses on the probability that a typical user triggers an HO in a unit of time. The LoS state of the user's serving link at the beginning of a typical unit of time is analyzed in this section.

\begin{figure*}[b]
\vspace{-0.2cm}
\hrulefill
\begin{equation} \label{PLink}
\setlength\abovedisplayskip{4pt}
\setlength\belowdisplayskip{4pt}
\begin{aligned}
&{\cal P}_{L,L}^{link}\left( {{x_1},{\phi _1}} \right) = \exp \left\{ { - \frac{{{\lambda _o}{\mathbb E}\left[ l \right]}}{\pi }\left( {2{x_2} - \!\!\!\!\!\!\!\! \int\limits_{\beta  \in \left[ {0,{\phi _1}} \right) \cup \left( {{\phi _2},\pi } \right)} \!\!\!\!\!\!\!\!{  {a\left( {1 - \frac{1}{2}a{{\dot a}_{\max }}} \right)}  {\rm{d}}\beta }  - \int_0^{{\phi _2}} {\vartheta {\rm{d}}\beta } } \right)} \right\},\ \ \ \ 
{\cal P}_{L,N}^{link}\left( {{x_1},{\phi _1}} \right) = 1 - {\cal P}_{L,L}^{link}\left( {{x_1},{\phi _1}} \right),
\\
&{\cal P}_{N,L}^{link}\left( {{x_1},{\phi _1}} \right) = \frac{{1 - \exp \left\{ { - \frac{{{\lambda _o}{\mathbb E}\left[ l \right]}}{\pi }\left( {2{x_1} -\!\!\!\!\!\!\!\! \int\limits_{\beta  \in \left[ {0,{\phi _1}} \right) \cup \left( {{\phi _2},\pi } \right)} \!\!\!\!\!\!\!\!{{a\left( {1 - \frac{1}{2}a{{\dot a}_{\max }}} \right)}{\rm{d}}\beta } } \right)} \right\}}}{{\left( {1 -  \exp \left\{ { - {\lambda _o}{\mathbb E}\left[ l \right]\frac{2}{\pi }{x_1}} \right\}} \right)\exp \left\{ {\frac{{{\lambda _o}{\mathbb E}\left[ l \right]}}{\pi }\left( {2{x_2} - \int_0^{{\phi _2}} {\vartheta {\rm{d}}\beta } } \right)} \right\}}},\ \ \ \ {\cal P}_{N,N}^{link}\left( {{x_1},{\phi _1}} \right) = 1 - {\cal P}_{N,L}^{link}\left( {{x_1},{\phi _1}} \right),
\\
&{{\dot a}_{\max }} = \left\{ \begin{array}{lr}
0,&\!\!\!{\phi _1},{\phi _2} \in \left\{ {0,\pi } \right\},\beta  = \frac{\pi }{2}\\
\frac{1}{{{a_{\max }}}},&\!\!\!{\rm others}
\end{array} \right.\!\!\!, \  a = \left\{ \begin{array}{lr}
{x_1},&\!\!\!\!\!\!\!\!\!\!\!\!\!\!\!\!\!\!\!\!\!\!\!\!\!\!\!\!\!\!\!\!\!\!\!\!\!\!\!\!\!\!\!\!\!\!\!{\phi _1},{\phi _2} \in \left\{ {0,\pi } \right\},\beta  = \frac{\pi }{2}\\
\min \left\{ {{x_1}\sin \left( {\left| {{\phi _1} - \beta } \right|} \right),{x_2}\sin \left( {\left| {{\phi _2} - \beta } \right|} \right),{a_{\max }}} \right\},&{\rm others}
\end{array} \right.\!\!,\ \ {a_{\max }} = \frac{{l\tan \varphi \tan \kappa }}{{\tan \varphi  + \tan \kappa }},\\ &\varphi  \!=\! \left\{ \begin{array}{ll}
\!\!\!\pi  \!-\! {\phi _2} \!+\! \beta ,\!\!\!\!&\!\beta  < {\phi _2}\\
\beta  \!-\! {\phi _2},\!&\beta  \ge {\phi _2}
\end{array} \right.\!\!\!,\ \ \kappa  \!=\! \left\{ \begin{array}{ll}
\!\!{\phi _1} \!-\! \beta ,& \!\!\beta < {\phi _1} \\
\!\!\pi  \!+\! {\phi _1} \!-\! \beta ,\!\!&\!\!\beta  \ge {\phi _1}
\end{array} \right.\!\!\!,\ \ 
\vartheta  \!=\! \left\{ \begin{array}{ll}
{\frac{{{\mathbb E}\left[ {{l^2}} \right]\sin \beta }}{{2{\mathbb E}\left[ l \right]}}\left( {\cos \beta  - \frac{{\sin \beta }}{{\tan {\phi _2}}}} \right),}&{l\cos \beta  - \frac{{l\sin \beta }}{{\tan {\phi _2}}} \le v}, {\phi _2} \notin \left\{ {0,\frac{\pi }{2},\pi } \right\}
\\
{v\sin \beta  - \frac{{{v^2}\sin \beta \tan {\phi _2}}}{{2{\mathbb E}\left[ l \right]\left( {\cos \beta \tan {\phi _2} - \sin \beta } \right)}},}&{l\cos \beta  - \frac{{l\sin \beta }}{{\tan {\phi _2}}} > v}, {\phi _2} \notin \left\{ {0,\frac{\pi }{2},\pi } \right\}
\\
v\sin \beta,& {\phi _2} \in \left\{ {0,\pi } \right\}
\\
\frac{{{\mathbb E}\left[ {{l^2}} \right]\sin \beta \cos \beta }}{{2{\mathbb E}\left[ l \right]}},&{\phi _2} = \frac{\pi }{2},l\cos \beta  \le v
\\
v\sin \beta  - \frac{{{v^2}\tan \beta }}{{2{\mathbb E}\left[ l \right]}},&{\phi _2} = \frac{\pi }{2},l\cos \beta  > v
\end{array} \right.\!\!\!\!.\\[-5mm]
\end{aligned}\tag{13}
\end{equation}
\vspace{-0.3cm}
\end{figure*}

According to {\it Lemma 1}, BSs with LoS states of LoS, NLoS, and reconfigured LoS for a typical user are distributed around the user at different densities. Owing to the principle that a user associates with the BS that provides the maximum received signal strength, the LoS state of the user's serving link at the beginning also has these three cases. The probability of each case is given by the following theorem:
\begin{theorem}
The association probabilities that the serving link of a typical user at the beginning of the unit of time is in the LoS state of LoS, NLoS, or reconfigured LoS are given in \eqref{Ak} (at the bottom of this page).
\end{theorem}

\begin{IEEEproof}
See Appendix C.
\end{IEEEproof}

\begin{remark}
From {\it Theorem 1}, it can be concluded that (i) ${\cal{A}}_L$ decreases when the ${{\mathbb E}\left[ l \right]}$ and $\lambda_o$ increase, partly because a decrease in $\lambda_{b,L}$ is followed by a increase in $\lambda_{b,N}$ and $\lambda_{b,R}$, and partly because a decrease in $\lambda_{b,L}$ is followed by a higher $d^L$ leading to a higher ${\widetilde d_L}$ and ${\widetilde r^L}$; (ii) ${\cal{A}}_R$ increases as $\lambda_i$ and $N$ increase, since higher $\lambda_i$ brings $\lambda_{b,R}$ raises and rightward shifts of ${\cal F}_{r_1}\left( {{r_1}\left| d \right.} \right)$, and higher $N$ brings higher ${\widetilde r^L}$ and ${\widetilde r^N}$, which is manifested by more users accessing NLoS BSs via IRSs, which provides a stronger signal, rather than accessing LoS BSs that are far away.
\end{remark}

\setcounter{equation}{10}

For the subsequent LoS state transition and HO analysis, the distribution of the distance between the user and its serving BS is given by the following theorem.
\begin{theorem}
For the serving link in LoS, NLoS, or reconfigured LoS, the cdf and pdf of the distance between a typical user and its serving BS are given by
\begin{equation}
\setlength\abovedisplayskip{7pt}
\setlength\belowdisplayskip{7pt}
\begin{aligned}
{{\cal F}_{x_1^k}}\left( {x_1^k} \right) &= 1 - \frac{{2\pi }}{{{{\cal A}_k}}}\int_{x_1^k}^\infty  {{\lambda _{b,k}}\left( {{d^k}} \right){d^k}{\Xi _k}\left( {{d^k}} \right){\rm{d}}{d^k}} ,\\
{f_{x_1^k}}\left( {x_1^k} \right) &= \frac{{2\pi }}{{{{\cal A}_k}}}{\lambda _{b,k}}\left( {x_1^k} \right)x_1^k{\Xi _k}\left( {x_1^k} \right), k \in \left\{ L, N, R \right\},
\end{aligned}
\end{equation}
where ${\Xi _k}\left( {d^k} \right)$ is given in \eqref{Fx0} (on bottom of this page) and ${\Xi _k}\left( {x_1^k} \right)$ is obtained by replacing ${d^k}$ with ${x_1^k}$ in ${\Xi _k}\left( {d^k} \right)$.
\end{theorem}

\begin{IEEEproof}
See Appendix D.
\end{IEEEproof}

\begin{remark}
For {\it Theorem 2}, the main discussion is on the distribution of $x_1^L$ (since the links corresponding to $x_1^R$ and $x_1^N$ have been blocked). When the blockage effect is exacerbated (i.e., higher $\lambda_o$ and ${{\mathbb E}\left[ l \right]}$), the cdf of $x_1^L$ is shifted right and $x_1^L$ is statistically larger, posing the risk of the link being blocked after movement. Larger $\lambda_i$ and $N$ can cause the cdf of $x_1^L$ to shift leftward, mainly due to a significant increase in $\int_0^\infty  {{\lambda _{b,R}}\left( {d^R} \right){{\cal F}_{{r_1}}}\left( {{{\widetilde r}^L}\left| {d^R} \right.} \right){d^R}{\rm{d}}{d^R}}$, which is manifested by the fact that users far away from the LoS BS are instead access NLoS BSs via IRSs.
\end{remark}

\begin{figure*}[b]
\vspace{-0.2cm}
\hrulefill
\begin{equation} \label{PBS}
\setlength\abovedisplayskip{4pt}
\setlength\belowdisplayskip{4pt}
\begin{aligned}
&{\cal P}_{L,k}^{BS} = \left\{ \begin{array}{ll}
{{\mathbb E}_{{x^L_1},{\phi _1}}}\left[ {{\cal P}_{L,L}^{link}\left( {{x^L_1},{\phi _1}} \right)} \right],&k = L\\
{{\mathbb E}_{x_1^L,{\phi _1}}}\left[ {{\cal P}_{L,N}^{link}\left( {x_1^L,{\phi _1}} \right)\left( {1 - {{\cal P}_{i}}\left( {{x_2^L},0} \right)} \right)} \right],&k = N\\
{{\mathbb E}_{x_1^L,{\phi _1}}}\left[ {{\cal P}_{L,N}^{link}\left( {x_1^L,{\phi _1}} \right){{\cal P}_{i}}\left( {{x_2^L},0} \right)} \right],&k = R
\end{array} \right.\!\!\!,\ \ 
{\cal P}_{N,k}^{BS} = \left\{ \begin{array}{ll}
{{\mathbb E}_{x_1^L,{\phi _1}}}\left[ {{\cal P}_{N,L}^{link}\left( {x_1^L,{\phi _1}} \right)} \right],&k = L\\
{{\mathbb E}_{x_1^N,{\phi _1}}}\left[ {{\cal P}_{N,N}^{link}\left( {x_1^L,{\phi _1}} \right)\left( {1 - {{\cal P}_{i}}\left( {x_2^N,0} \right)} \right)} \right],&k = N\\
{{\mathbb E}_{x_1^N,{\phi _1}}}\left[ {{\cal P}_{N,N}^{link}\left( {x_1^L,{\phi _1}} \right){{\cal P}_{i}}\left( {x_2^N,0} \right)} \right],&k = R
\end{array} \right.\!\!\!,
\\
&{\cal P}_{R,k}^{BS} = \left\{ \begin{array}{ll}
{{\mathbb E}_{x_1^R,{\phi _1}}}\left[ {{\cal P}_{N,L}^{link}\left( {x_1^R,{\phi _1}} \right)} \right],&k = L\\
{{\mathbb E}_{x_1^R,{\phi _1}}}\left[ {{\cal P}_{N,N}^{link}\left( {x_1^R,{\phi _1}} \right)} \right] \times {{\mathbb E}_{{r_1},{\phi '_1},x_1^R}}\left[ {{\cal P}_{L,N}^{link}\left( {{r_1},{\phi '_1}} \right) \times \left( {1 - {{\cal P}_{i}}\left( {x_2^R,{r_2}} \right)} \right)} \right],&k = N\\
{{\mathbb E}_{x_1^R,{\phi _1}}}\left[ {{\cal P}_{N,N}^{link}\left( {x_1^R,{\phi _1}} \right)} \right] \times \left( {1 - {{\mathbb E}_{{r_1},{\phi '_1},x_1^R}}\left[ {{\cal P}_{L,N}^{link}\left( {{r_1},{\phi '_1}} \right) \times \left( {1 - {{\cal P}_{i}}\left( {x_2^R,{r_2}} \right)} \right)} \right]} \right),&k = R
\end{array} \right..\\[-2mm]
\end{aligned}\tag{14}
\end{equation}
\begin{equation} \label{HOP}
\begin{aligned}
&{\cal H} = \sum\limits_{k \in \left\{ {L,N,R} \right\}} {{{\cal A}_k}\left\{ {\sum\limits_{j \in \left\{ {L,N,R} \right\}} \!\!{{\cal P}_{k,j}^{BS} \left( 1-\!\!\! \prod\limits_{w \in \left\{ {L,N} \right\}} {{{\mathbb E}_{x_1^k,{{\phi}_1}}}\left[ {{\bar{\cal H}^{k,j,w}}} \right]} \right) } } \right\}} ,\ \ \ {\bar{\cal H}^{k,j,w}} = \left\{ \begin{array}{ll}
 \exp \left\{ {{\Omega ^{k,j,w}}} \right\},&k,j \in \left\{ {L,N} \right\}\\
 {{\mathbb E}_{{r_1}}}\left[ {\exp \left\{ {{\Omega ^{k,j,w}}} \right\}} \right],&
k = R,j \in \left\{ {L,N} \right\}\\&{ \!\!\!\!\!\!\!\cup }\ k \in \left\{ {L,N} \right\},j = R\\
 {{\mathbb E}_{{r_1},{\phi '_1}}}\left[ {\exp \left\{ {{\Omega ^{k,j,w}}} \right\}} \right],&k = j = R
\end{array} \right.,\\
&{\Omega ^{k,j,w}} =  - \int_{\max \left\{ {0,x_{1,eq}^{k,w} - v} \right\}}^{x_{2,eq}^{j,w}} {\int_{ - {\theta _{\max }}}^{{\theta _{\max }}} {{\lambda _{b,w}}\left( x \right)x{\rm{d}}\theta {\rm{d}}x} }, \ \  {\theta _{\max }} = \left\{ \begin{array}{ll}
\pi ,&0 < x \le v - x_{1,eq}^{k,w},v \ge x_{1,eq}^{k,w}\\
\arccos \frac{{{{\left( {x_1^{k,q}} \right)}^2} - {v^2} - {x^2}}}{{2vx}},&v - x_{1,eq}^{k,w} < x < x_{2,eq}^{j,w}
\end{array} \right.,
\\
&x_{c,eq}^{L,w} = \left\{ \begin{array}{ll}
x_c^L,&\!\!\!\!\!w = L\\
{\left( {\frac{{{K_N}}}{{{K_L}}}} \right)^{\frac{1}{{{\alpha _N}}}}}{\left( {x_c^L} \right)^{\frac{{{\alpha _L}}}{{{\alpha _N}}}}},&\!\!\!\!\!w = N
\end{array} \right.\!\!\!,\ x_{c,eq}^{N,w} = \left\{ \begin{array}{ll}
{\left( {\frac{{{K_L}}}{{{K_N}}}} \right)^{\frac{1}{{{\alpha _L}}}}}{\left( {x_c^N} \right)^{\frac{{{\alpha _N}}}{{{\alpha _L}}}}},&\!\!\!\!\!w = L\\
{x^N_c},&\!\!\!\!\!w = N
\end{array} \right.\!\!\!,\ x_{c,eq}^{R,w} = \left\{ \begin{array}{ll}
{\left( {\frac{1}{{{K_L}{G_{bf}}}}} \right)^{\frac{1}{{{\alpha _L}}}}}{r_c}x_c^R,&\!\!\!\!\!w = L\\
{\left( {\frac{{{K_N}}}{{K_L^2{G_{bf}}}}} \right)^{\frac{1}{{{\alpha _N}}}}}{\left( {{r_c}x_c^R} \right)^{\frac{{{\alpha _L}}}{{{\alpha _N}}}}},&\!\!\!\!\!w = N
\end{array} \right.\!\!\!, c \in \left\{1,2 \right\}.\\[-4mm]
\end{aligned} \tag{16}
\end{equation}
\end{figure*}

\subsection{LoS State Transitions}

Before analyzing whether the user will be handed over to a new BS, the transition of the LoS state of the link between the user and the original BS due to user movment is analyzed in this section.
We first consider LoS state transitions of the single link of user-BS or user-IRS, where the transitions only involve the states of LoS and NLoS.

Without any loss of generality, we consider the location of the typical user at the beginning of the unit of time as the origin and its direction of movement in the typical unit of time as the $x$-axis positive direction. The angle between the user-BS/IRS line at the initial location (or after movement) and the direction of movement is denoted by ${\phi _1}$ (${\phi _2}$), and the distance between the typical user and the BS/IRS is denoted by ${x _1}$ (${x _2}$). We have ${\phi _2} = \arccos \left( {\frac{{{x_1}\cos {\phi _1} - v}}{{{x_2}}}} \right)$, ${x_2} = \sqrt {x_1^2 + {v^2} - 2{x_1}v\cos {\phi _1}}$.
Then, the transition probabilities of a single link are given by the following lemma: 
\begin{lemma}
Regarding the distance of the single link $x_1$ and the angle relative to the direction of movement ${\phi _1}$, the transition probabilities of LoS states of a single link are given in \eqref{PLink} (at the bottom of this page), where the first and second items of the footprint in ${\cal P}_{k,j}^{link},k,j \in \left\{ L, N \right\}$ represent the  LoS state of the user before and after unit time, respectively.
\end{lemma}

\begin{IEEEproof}
See Appendix E.
\end{IEEEproof}

\begin{remark}
For {\it Lemma 3}, the main discussion is on the probability of the LoS link being blocked after movement (i.e., ${\cal{P}}^{link}_{L,N}\left( {{x_1},{\phi _1}} \right)$), since it is the key to causing frequent HOs (whether it happens on the BS-user link or the IRS-user link). From \eqref{PLink}, it can be conclude that (i) ${\cal{P}}^{link}_{L,N}\left( {{x_1},{\phi _1}} \right)$ is mainly determined by the expectation of two areas, one is ${\mathbb E}\left[ {\left| {{{\cal S}^{{t_2}}}} \right|} \right] = \frac{2}{\pi }{\mathbb E}\left[ l \right]{x_2}$ corresponding to possible locations of blockage midpoints when the link is blocked after the move, and one is ${\mathbb E}\left[ {\left| {{\cal{S}}_ \cap ^{{t_1},{t_2}}} \right|} \right] = \frac{2}{\pi }{\mathbb E}\left[ l \right]\int_{\beta  \in \left[ {0,{\phi _1}} \right) \cup \left( {{\phi _2},\pi } \right)} {a\left( {1 - \frac{1}{2}a{{\dot a}_{\max }}} \right){\rm{d}}\beta } $ corresponding to possible locations of blockage midpoints when the link is blocked both before and after the move, so a smaller $x_1$ implies a smaller ${\cal{P}}^{link}_{L,N}\left( {{x_1},{\phi _1}} \right)$, but given ${x_1}$, a larger $\phi_1$ which causes the user to move farther away from the BS (i.e., a larger $x_2$), may not imply a larger ${\cal{P}}^{link}_{L,N}\left( {{x_1},{\phi _1}} \right)$, since ${\mathbb E}\left[ {\left| {{{\cal S}^{{t_2}}}} \right|} \right] - {\mathbb E}\left[ {\left| {{{\cal S}}_ \cap ^{{t_1},{t_2}}} \right|} \right]$ increases and then decreases with ${\phi _1}$; (ii) ${\cal{P}}^{link}_{L,N}\left( {{x_1},{\phi _1}} \right)$ increases with ${{\mathbb E}\left[ l \right]}$ and $\lambda _i $ and asymptotically approaches 1. 
\end{remark}

\setcounter{equation}{14}

Based on the analysis of the single link, we further consider the cascaded link of the user, BS, and IRS, whose transition involves states of LoS, NLoS, and reconfigured LoS. The transition probabilities of the cascaded link are then given by the following theorem:

\begin{theorem}
The transition probabilities of LoS states between the user and serving BS in IRS-aided networks are provided in \eqref{PBS} (at the bottom of next page), where 
\begin{equation}
\setlength\abovedisplayskip{1pt}
\setlength\belowdisplayskip{1pt}
\begin{aligned}
{{\cal P}_i}\left( {d,r} \right) &= 1 - {e^{ - {\lambda _i}\int\limits_{ - \pi }^\pi  {\int\limits_r^D {{p_i}\left( {r,\theta \left| d \right.} \right)r{\rm{d}}r{\rm{d}}\theta } } }}\\
 &  \approx  1 - {e^{ - {\lambda _i}{{\widehat {p}}_i}\left( {r,D,d} \right)}} 
\end{aligned}
\end{equation}
is the probability of existing available IRSs outside the distance $r$ from the user when user-BS distance is $d$, ${\phi _1}$ and ${\phi '_1}$ are both uniformly distributed in $\left[ { - \pi ,\pi } \right)$, $x_2^k = \sqrt {{{\left( {x_1^k} \right)}^2} + {v^2} - 2x_1^kv\cos {\phi _1}} ,k \in \left\{ {L,N,R} \right\}$, ${r_2} = \sqrt {{{\left( {{r_1}} \right)}^2} + {v^2} - 2{r_1}v\cos {\phi '_1}} $.
\end{theorem}

\begin{IEEEproof}
See Appendix F.
\end{IEEEproof}

\begin{remark}
In {\it Theorem 3}, we discuss the case where the serving link falls from LoS or reconfigured LoS to NLoS, which with high probability causes HOs. For LoS links, maintaining LoS can be achieved by increasing the BS density and thus reducing the $x_1^L$ and increasing ${\cal P}_{L,L}^{link}\left( {x_1^L,{\phi _1}} \right)$. From ${\cal P}_{L,R}^{BS}\left( {x_1^L,{\phi _1}} \right)$, IRS offers the possibility of building reconfigured LoS links when the BS-user link is blocked due to movement,  thus, it is crucial to enhance ${{\cal P}_i}\left( {x_2^L,0} \right)$ (related discussions are given in {\it Remark 1}). For reconfigured LoS links, a smaller $r_1$ facilitates the maintenance of the LoS with the original IRS (discussions on $r_1$ are given in {\it Remark 2}), and ${{\cal P}_i}\left( {x_2^L,r_2} \right)$ still keeps the possibility of finding a new IRS to avoid dropping into NLoS. Therefore, the IRS provides a novel way to avoid frequent HOs.
\end{remark}

\begin{figure}[t]
\centerline{\includegraphics[width=0.8\linewidth]{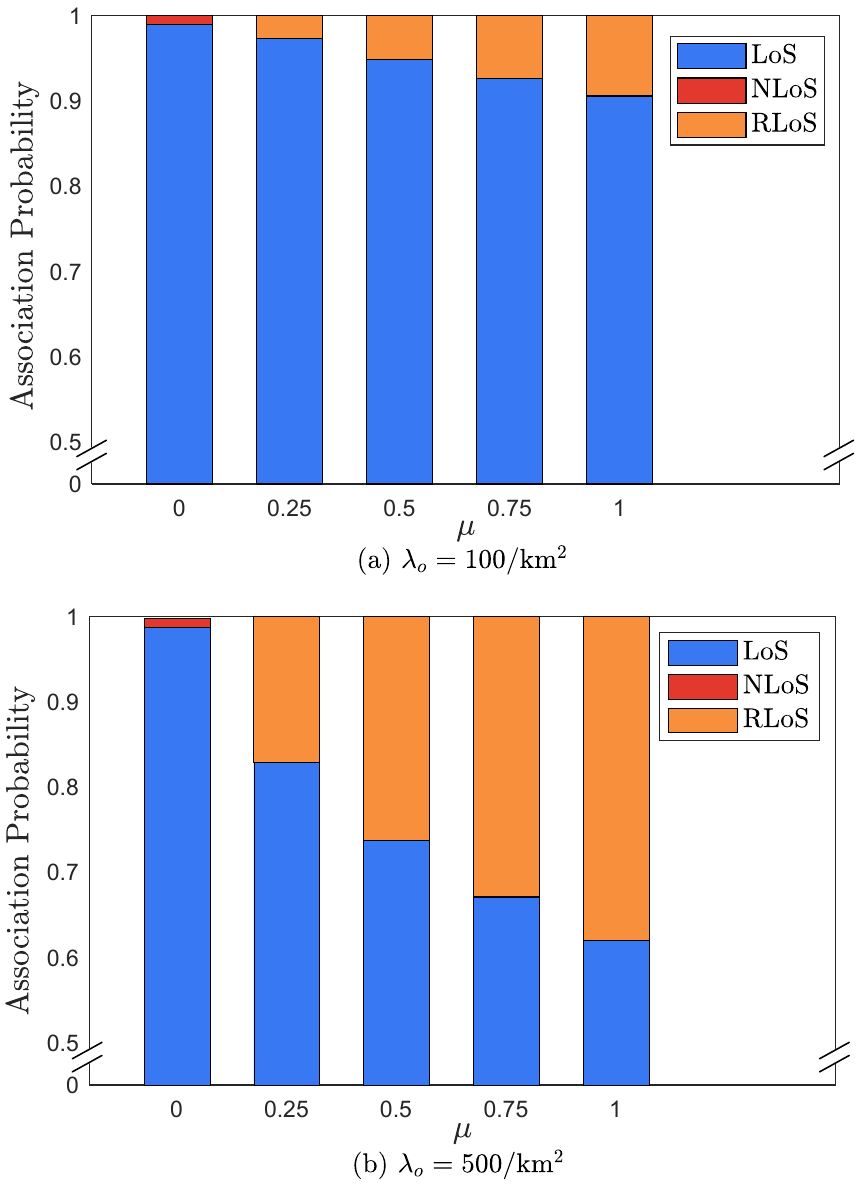}}
\vspace{-0.3cm}
\caption{Association probabilities of serving link at the beginning of the unit time as functions of the percentage of blockages that are equipped with IRSs $\mu$ under different blockage densities $\lambda_o$: (a) $\lambda_o=100 / {\rm km^2}$; (b) $\lambda_o=500 / {\rm km^2}$.}
\vspace{-0.3cm}
\label{fig1}
\end{figure}

\begin{figure*}[t]
\centerline{\includegraphics[width=0.95\linewidth]{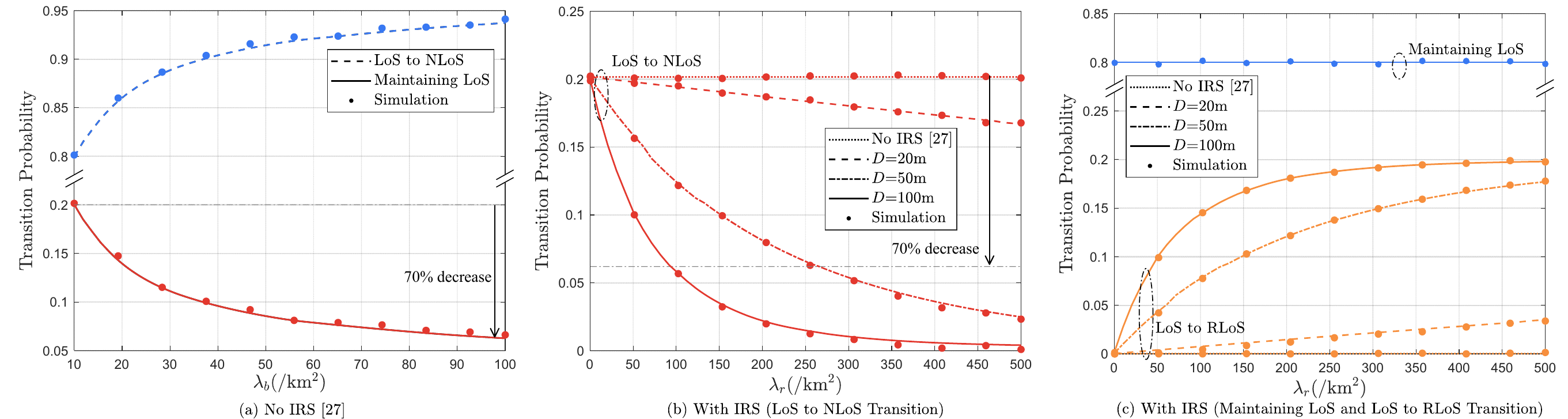}}
\vspace{-0.3cm}
\caption{Transition probabilities of LoS states of the serving link in networks with/without IRS aiding: (a) no IRS [27]; (b) with IRS (LoS to NLoS transition); (c) with IRS (maintaining LoS and LoS to RLoS transition).}
\end{figure*}

\begin{figure}[t]
\centerline{\includegraphics[width=0.85\linewidth]{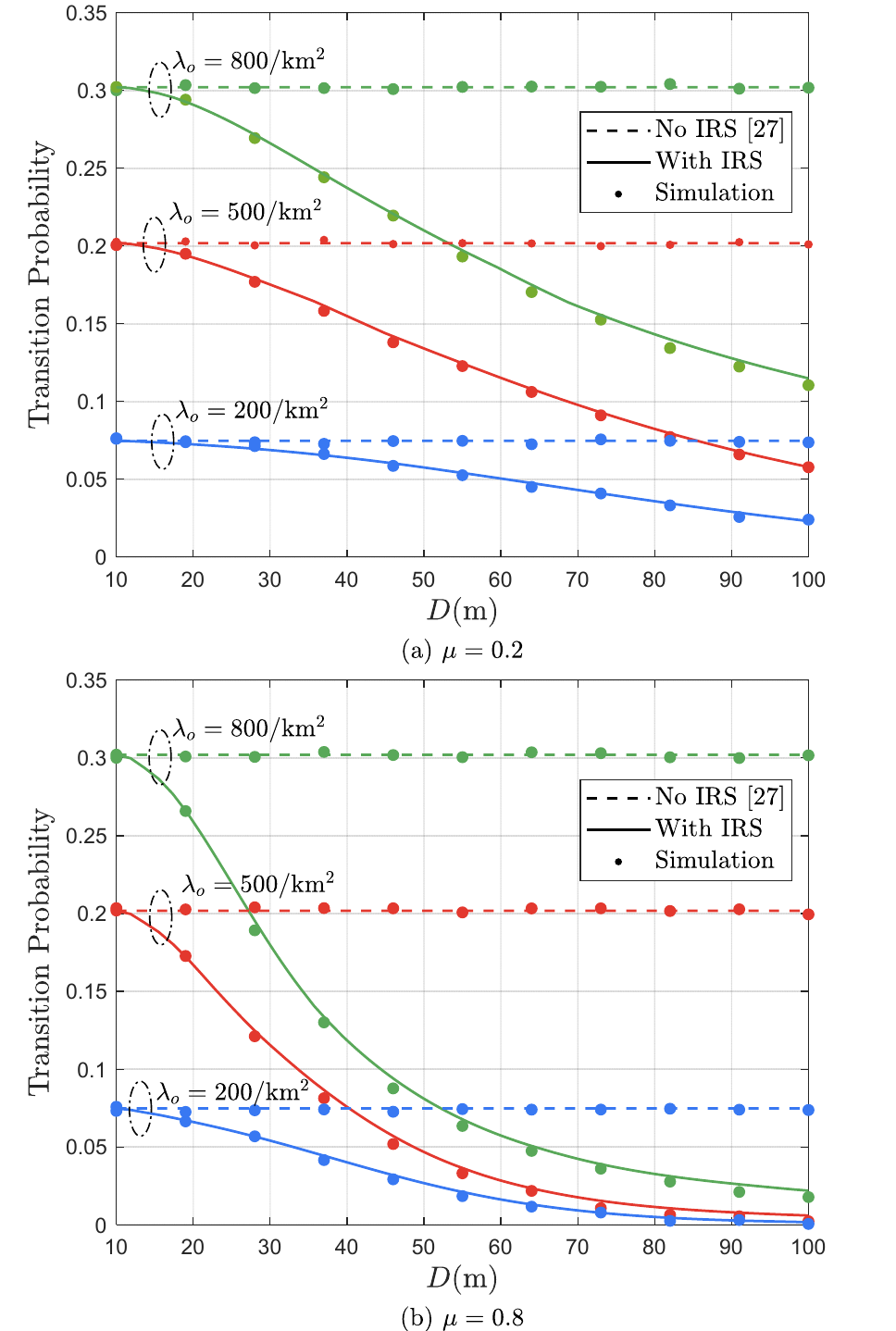}}
\vspace{-0.3cm}
\caption{Transition probabilities of LoS states of the serving link (LoS to NLoS) in networks with/without IRS as functions of IRS serving distance $D$ under different blockage densities $\lambda_o$ and percentages of blockages that are equipped with IRSs $\mu$: (a) $\mu = 0.2$; (b) $\mu = 0.8$.}
\vspace{-0.2cm}
\label{fig1}
\end{figure}

\begin{figure}[t]
\centerline{\includegraphics[width=0.85\linewidth]{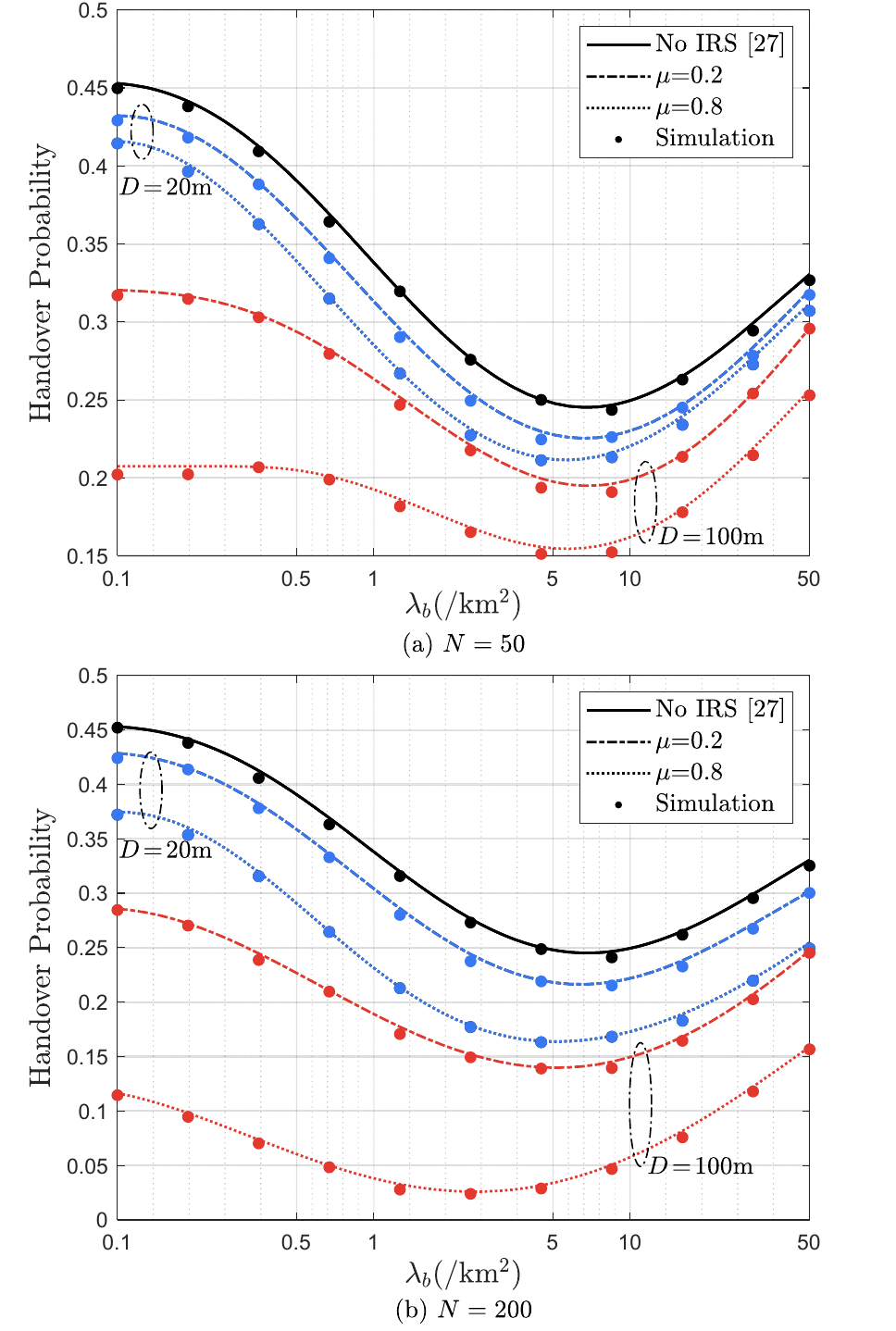}}
\vspace{-0.3cm}
\caption{HO probabilities in networks with/without IRS as functions of BS density $\lambda_b$ under different IRS serving distances $D$ and numbers of IRS elements $N$: (a) $N = 50$; (b) $N = 200$.}
\vspace{-0.2cm}
\label{fig1}
\end{figure}

\vspace{-0.2cm}
\subsection{Handover Probability}

Based on the analysis of the initial LoS states and transitions of LoS states in IRS-aided networks, the decision on whether HO is triggered is analyzed in this section. The HO probability is given by the following theorem:
\begin{theorem}
The HO probability (denoted as $\mathcal{H}$) of a typical user in IRS-aided networks considering blockage effects is presented in \eqref{HOP} (at the bottom of this page).
\end{theorem}

\begin{IEEEproof}
See Appendix G.
\end{IEEEproof}

\begin{remark}
From {\it Theorem 4}, it can be concluded that (i) the HO probability increases significantly when the link falls into NLoS due to $x_{2,eq}^{N,L} \!\!\gg\!\! x_{2,eq}^{L,L}$, thus reducing the probabilities ${\cal P}^{BS}_{L,N}$ and ${\cal P}^{BS}_{R,N}$ is crucial (related discussions are given in {\it Remark 6}); (ii) the signal strength from the IRS also determines the effect of avoiding unnecessary HOs, which is manifested at $x_{2,eq}^{R,L} \!\!\propto\!\! \frac{{{r_c}}}{{\sqrt[{{\alpha _L} \ }]{{{G_{\!bf}}}}}}$, thus, only if the IRS is dense enough and $N$ is large enough can the HOs be practically reduced, otherwise the advantages of IRS cannot be exploited.
\end{remark}

%\begin{figure}[t]
%\centerline{\includegraphics[width=0.8\linewidth]{figfinal0/fig2b.eps}}
%\vspace{0.1cm}
%\centerline{\includegraphics[width=0.8\linewidth]{figfinal0/fig2c.eps}}
%\vspace{-0.4cm}
%\caption{Transition probabilities of LoS states of serving link in IRS-aided networks as functions of IRS density $\lambda_r$: (a) LoS to NLoS transition; (b) Maintaining LoS, and LoS to reconfigured LoS transition.}
%\vspace{-0.3cm}
%\label{fig1}
%\end{figure}

\section{Numerical Results}

According to \cite{b22,b25,sp1}, the following parameters are adopted if not specific: $p_{t}=24\rm{dBm}$, ${K _L}=10^{-10.38}$, ${K_N}=10^{-14.54}$, ${\alpha _L}=2.09$, ${\alpha _N}=3.75$, $m_L = 10$, $m_N = 1$,  $\lambda_b=10/\rm{km^2}$, $\lambda_o=500/\rm{km^2}$, $\mu=0.5$, $l=10{\rm{m}}$ (set as a constant), $N=500$, $D={\rm 50m}$, $v=20\rm{m}/\rm{s}$. Monte Carlo simulations are carried out to validate our analysis. The reconfigured LoS is abbreviated as RLoS in the figures in this section. For comparison, the results for the case of no IRS (as in \cite{b21}) are also presented. %\cite{b21}

Fig. 3 illustrates the association probabilities of the initial LoS state being in LoS, NLoS, or reconfigured LoS as functions of $\mu$ (i.e., the percentage of blockages equipped with IRSs) under different blockage densities $\lambda_o$, where theoretical results are plotted via Eq. \eqref{Ak} and $\mu=0$ corresponds to the case of no IRS [27]. In the absence of IRS deployment ($\mu=0$), a small fraction of the users ($<2\%$) initially experience NLoS conditions. However, with the introduction of IRSs, the users in the NLoS condition disappear. As $\mu$ increases, more serving links are observed to be in the reconfigured LoS at the initial location. This results from the reconfigured LoS links by IRSs, providing enhanced received signals compared with direct LoS. This trend becomes more pronounced for high-density blockages $(\lambda_o=500/\rm{km^2})$. Specifically, when $\lambda_o=500/\rm{km^2}$ and $\mu=1$, approximately 38\% of users initially experience reconfigured LoS links.

\begin{figure}[t]
\centerline{\includegraphics[width=0.85\linewidth]{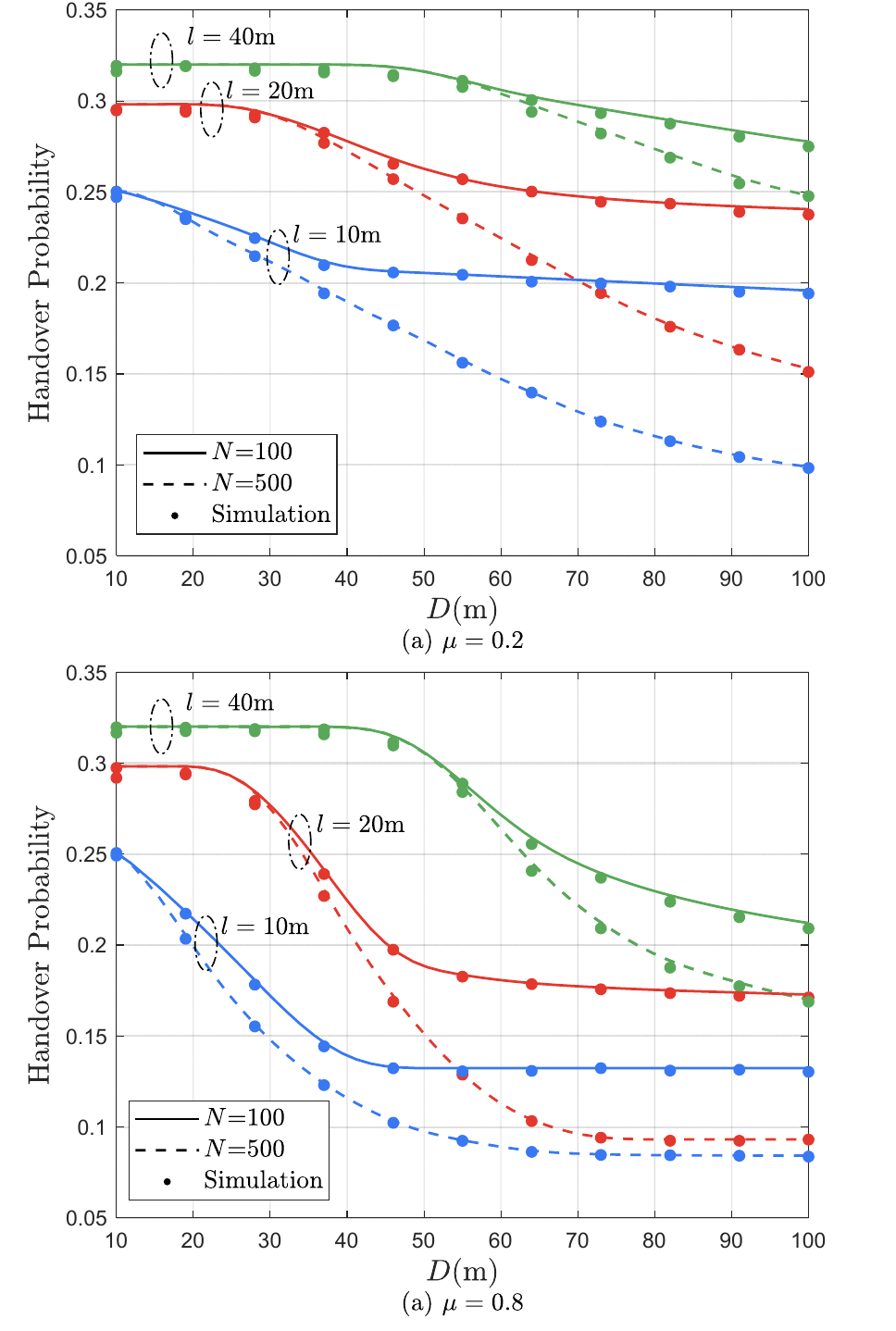}}
\vspace{-0.3cm}
\caption{HO probabilities in IRS-aided networks as functions of IRS serving distance $D$ under different numbers of IRS elements $N$ and percentages of blockages that are equipped with IRSs $\mu$: (a) $\mu = 0.2$; (b) $\mu = 0.8$.}
\vspace{-0.2cm}
\label{fig1}
\end{figure}

\begin{figure}[t]
\centerline{\includegraphics[width=0.85\linewidth]{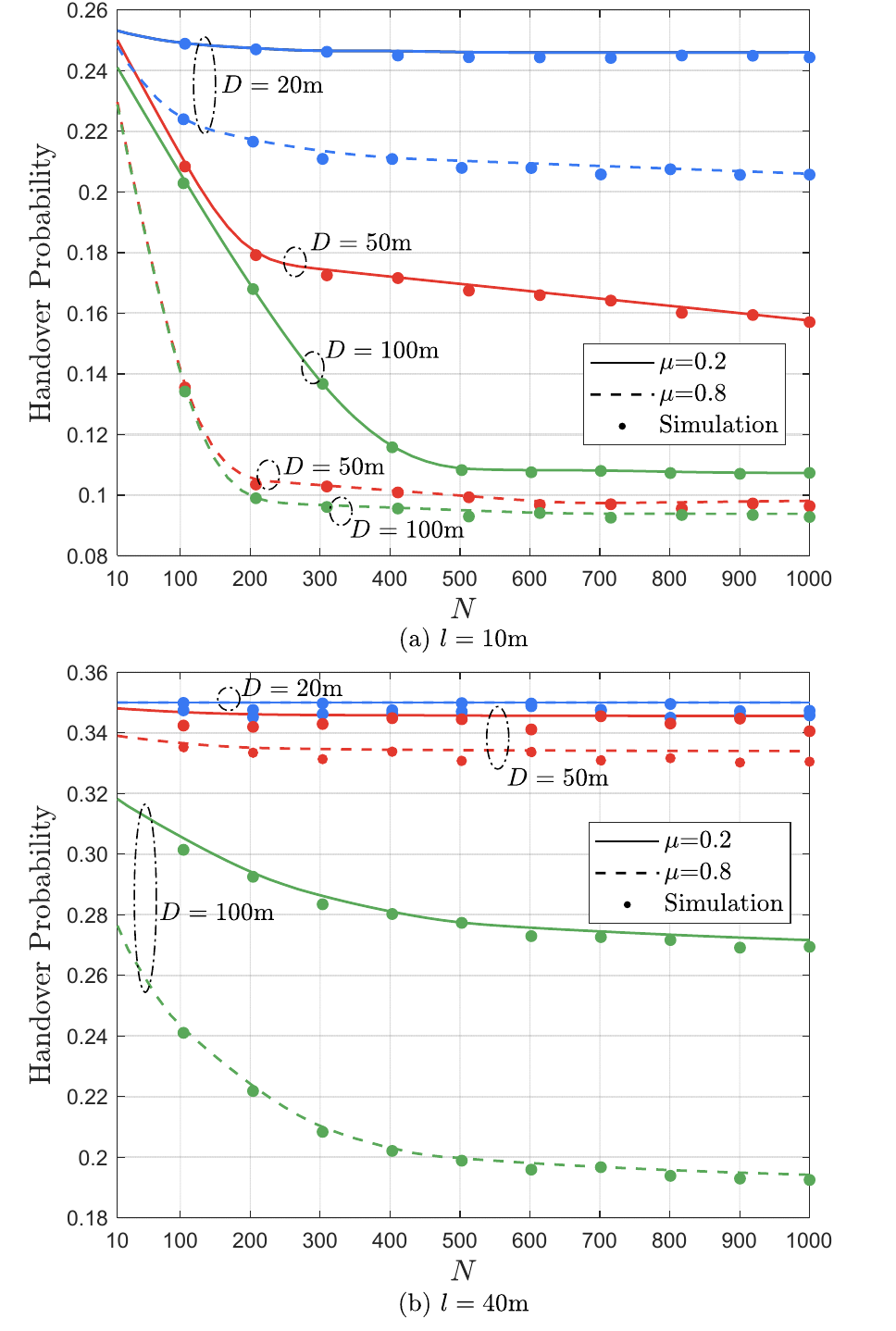}}
\vspace{-0.3cm}
\caption{HO probabilities in IRS-aided networks as functions of the number of IRS elements $N$ under different IRS serving distances $D$ and blockage lengths $l$: (a) $l = 10 {\rm m}$; (b) $l = 40 {\rm m}$.}
\vspace{-0.2cm}
\label{fig1}
\end{figure}

\begin{figure*}[!t]
\centerline{\includegraphics[width=480pt]{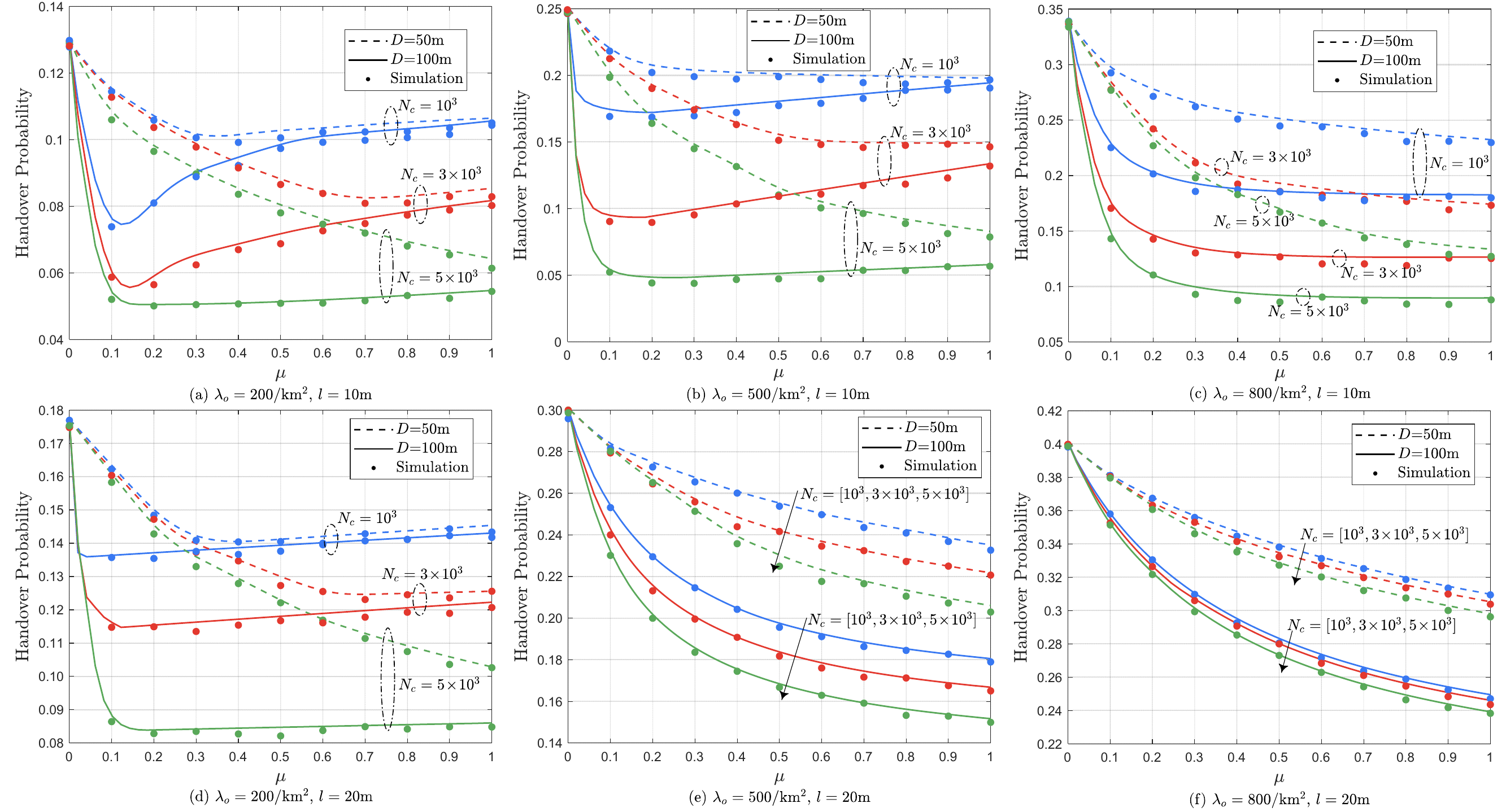}}
\vspace{-0.3cm}
\caption{HO probabilities as functions of the percentage of blockages that are equipped with IRSs $\mu$ with the fixed total number of IRS elements per cell ${N_{c}} \!=\! \frac{{{\lambda _r}}}{{{\lambda _b}}}N$ under different blockage densities $\lambda_o$ and blockage lengths $l$:
(a) $\lambda_o=200/{\rm km^2}$, $l=10 {\rm m}$;
(b) $\lambda_o=500/{\rm km^2}$, $l=10 {\rm m}$;
(c) $\lambda_o=800/{\rm km^2}$, $l=10 {\rm m}$;
(d) $\lambda_o=200/{\rm km^2}$, $l=20 {\rm m}$;
(e) $\lambda_o=500/{\rm km^2}$, $l=20 {\rm m}$;
(f) $\lambda_o=800/{\rm km^2}$, $l=20 {\rm m}$.}
\label{fig1}
\vspace{-0.4cm}
\end{figure*}

Fig. 4(a) demonstrates the impact of increasing BS density as an approach to ameliorate the occurrence of LoS links transitioning into NLoS links owing to user mobility, where theoretical results are plotted via Eq. \eqref{PBS}. 
 When the serving link falls into NLoS, the received signal strength experiences significant attenuation and frequent intracell HOs occur, resulting in additional network overhead and power consumption at the terminals, among others. Specifically, as the BS density increases from $10/\rm{km^2}$ to $100/\rm{km^2}$, the probability of transition to NLoS decreases from 0.2 to 0.06, representing a notable 70\% reduction. While this approach shows improvement in mitigating NLoS occurrences, it comes at the cost of a ten-fold increase in BS density, introducing significant infrastructure deployment costs.

Figs. 4(b) and 4(c) illustrate the impact of increasing the IRS density $\lambda_i$ and extending the IRS serving distance $D$ to mitigate the occurrence of LoS links transitioning into NLoS, where theoretical results are plotted via Eq. \eqref{PBS}.
 As the IRS density and serving distance increase, more serving links that would otherwise fall into NLoS are reconfigured by the IRS. The reconfigured LoS link proves beneficial for avoiding signal fluctuations and extra HOs resulting from the LoS-to-NLoS transition. Specifically, when the IRS density and serving distance reach $93/\rm{km^2}$ and $100\rm{m}$, respectively, the probability of transitioning to NLoS decreases to 0.06. This enhancement is comparable to the effect of increasing BS density 10 times as shown in Fig. 4(a). Considering the lower cost and energy consumption associated with IRS deployment, the results demonstrate the effectiveness of deploying IRSs as an efficient approach to prevent LoS links from transitioning into NLoS.

Fig. 5 shows the impact of different blockage densities $\lambda_o$ on the effectiveness of IRSs in mitigating the transition of serving links to NLoS owing to user mobility, where theoretical results are plotted via Eq. \eqref{PBS}. 
 With a high blockage density, deploying IRSs continues to be effective in reducing the probability of serving links falling into NLoS. However, this effectiveness is influenced by IRS density and serving distance. Specifically, at a lower IRS density ($\mu=0.2$), even with the extended IRS serving distance of $100 {\rm m}$, over 10\% of the serving links still transition to NLoS. With sufficient IRS density ($\mu=0.8$), an IRS serving the distance of $38 {\rm m}$ is adequate to reduce the probability to below 10\%. While IRSs prove effective in preventing serving links from transitioning to NLoS, the issue of whether or not IRSs can provide sufficient gain of the received signal strength to avoid unnecessary HOs remains to be further discussed.

%\cite{b11, b13}
In Fig. 6, the HO probabilities are plotted as a function of BS density $\lambda_b$ under different percentages of blockages equipped with IRSs $\mu$, numbers of IRS elements $N$, and IRS serving distances $D$, where theoretical results are plotted via Eq. \eqref{HOP}. In contrast to previous works, such as \cite{b11, b13} etc., which ignored blockage effects and reported a monotonically increasing trend in the HO probability with increasing BS density, the results of this paper reveal a new trend: The HO probability initially decreases and then increases with increasing BS density. This is due to the positive effect of the increased BS density, which reduces the user-to-BS distance, and thus the number of LoS links transitioning to NLoS. However, further increasing the BS density introduces more candidate BS, leading to an increase in HOs. Fig. 8 shows the existence of an optimal BS density configuration that minimizes the probability of HOs. Furthermore, the introduction of IRSs results in lower minimum HO probabilities and optimal BS densities. Specifically, with $N=50$, $\mu=0.2$ and $D=20\rm{m}$, the optimal $\lambda_b$ is $6.7/\rm{km^2}$ with the HO probability of 0.22. For $N=200$, $\mu=0.8$ and $D=100\rm{m}$ the optimal $\lambda_b$ is $2.8/\rm{km^2}$ with the HO probability of 0.028.

Fig. 7 presents insightful observations regarding the impact of IRS serving distance $D$ on the HO probability under varying blockage lengths $l$, numbers of IRS elements $N$, and percentages of blockages equipped with IRSs $\mu$, where theoretical results are plotted via Eq. \eqref{HOP}. When both the number of IRS elements and the percentage of blockages equipped with IRSs are relatively low ($N=100$, $\mu=0.2$), increasing the IRS serving distance has a limited effect on reducing HOs. Specifically, at $l$ values of $40{\rm m}$ and $10{\rm m}$, the maximum reductions in the HO probability are 4.2\%  and 5.5\%, respectively. This constrained improvement is attributed to the larger $D$, which enhances the probability of reconfiguring LoS links; however, the increased user IRS distance and fewer IRS elements fail to provide sufficient channel gain, resulting in unavoidable HOs. Notably, when $\mu=0.8$, a nearly constant trend of the HO probability with $D$ emerges, and as $N$ increases, the HO probability sustains a more consistent decrease.

Fig. 8 shows the interplay between the number of IRS elements $N$, blockage length $l$, IRS serving distance $D$, and percentage of blockages equipped with IRSs $\mu$ influencing the HO probability, where theoretical results are plotted via Eq. \eqref{HOP}. The results show that increasing $N$ is advantageous for ensuring a strong received signal when the IRS reconfigures LoS links, thereby preventing unnecessary HOs. However, this enhancement in $N$ does not affect the probability of reconfiguring LoS links. Consequently, when the probability of reconfiguring LoS links is low, the impact of increasing $N$ is limited. For instance, when $\mu=0.2$, $D=100{\rm m}$, and $N$ increases from $10$ to $1000$, the reductions in HO probability are 13.4\% and 4.7\% for blockage lengths $l$ of $10{\rm m}$ and $40{\rm m}$, respectively. Figs. 7 and 8 collectively underscore that effectively reducing the HO probability requires simultaneous improvements in the IRS configurations: number of elements, density, and serving distance. Moreover, the figures emphasize that once a certain parameter is enhanced to a certain extent, further improvements yield diminishing returns.

Fig. 9 provides comprehensive insights into the impact of the distributed deployment of IRSs on the HO probability under various scenarios characterized by the IRS serving distance $D$, total number of IRS elements per cell $N_c$, blockage length $l$, and blockage density $\lambda_o$, where theoretical results are plotted via Eq. \eqref{HOP} and $\mu=0$ corresponds to the case of no IRS [27]. Notably, this study considers a fixed total number of IRS elements per cell, denoted as $N_c$; therefore, we have ${N_{c}} \!=\! \frac{{{\lambda _r}}}{{{\lambda _b}}}N$. When the number of elements per IRS $N$, the IRS density decreases accordingly. The results reveal trends influenced by factors such as blockage density $\lambda_o$ and length $l$. In cases with a low blockage density and length, there exists an optimal $\mu^*$ that minimizes the HO probability. For instance, when $\lambda_o=200/\rm{km^2}$, $l=10{\rm m}$, and $N_c = 3 \times 10^3$ with $D=50{\rm m}$ and $D=100{\rm m}$, $\mu^*$ is 0.7 and 0.14, resulting in HO probabilities of 8.2\% and  5.6\% (37\% and 57\% reduction compared to [27]), respectively. Increasing $N_c$ effectively reduces the HO probability, as shown in Fig. 11(d). In contrast, in cases with a high blockage density and length, as shown in Figs.11(c), (e), and (f), the HO probability monotonically decreases with increasing $\mu$. Here, the probability of reconfiguring LoS links significantly influences the HO probability. In these cases, the most distributed IRS configuration leads to the lowest HO probability, even at the cost of reduced IRS element numbers $N$ and reconfigured LoS link gain. The impact of increasing $N_c$ on the reduction in the HO probability is constrained, as illustrated in Fig. 11(f), where an increase in $N_c$ from $10^3$ to $5 \times 10^3$ with $D=50 {\rm m}$ only results in a relative decrease of 4.7\% in the HO probability. These findings offer valuable guidance for optimizing the deployment of IRSs.

\vspace{0.2cm}

\section{Conclusion}

This paper proposes an analytical HO model for IRS-aided networks to explore the mobility enhancement achieved by reconfigured LoS links through IRS, which includes an exact analysis of IRS-reconfigured LoS links, transitions of LoS states, and HO decisions. The theoretical results of the probabilities of LoS state transitions and HO reveal valuable design insights which are summarized as: (i) deploying IRSs is proven to be an efficient way to prevent the transition of LoS links to NLoS due to user mobility, e.g., the probability of dropping into NLoS due to user movement decreases by 70\% when deploying IRSs with the density of $\rm{ 93/km^2}$, which is equivalent to the effect of increasing BS density from $\rm{ 10/km^2}$ to $\rm{ 100/km^2}$; (ii) there is a tradeoff between avoiding falling into NLoS and frequent HOs when raising BS density, where optimal BS density exists and is affected by IRS configuration; (iii) limitations in any one of the IRS parameters (including: density, number of elements, serving distance) restrict the IRS gain, thus  the parameters need to be raised comprehensively; (iv) optimal IRS distributed deployment parameter exists that minimizes the HO probability, and more distributed IRSs are superior as the blockage effects are severer.

\vspace{-0.2cm}
\begin{appendix}
\vspace{-0.2cm}

\subsection{Proof of Lemma 1}
\vspace{-0.1cm}

As shown in Fig. 10, for a given ${\beta }$ and $l$, the propagation path from the user to the BS is blocked if and only if at least one blockage exists whose midpoint falls in the quadrilateral AEFD, which is denoted as ${{\cal D}_1}$. Similarly, for the user-IRS path and IRS-BS path, the quadrilaterals BCFE and ABCD are considered, which are denoted as ${{\cal R}_1}$ and ${\cal D}'$, respectively.

\setcounter{equation}{16}

Hence, the LoS link between the BS and the user implies that there is no blockage with the midpoint falling into area ${{\cal D}_1}$. The probability of an LoS link between the user and the BS at distance $d$ is given by \cite{b6}
\begin{equation}
\setlength\abovedisplayskip{3pt}
\setlength\belowdisplayskip{3pt}
\begin{aligned}
{p_{L}}\left( d \right) &= {e^{ - {\lambda _o}{{\mathbb E}_{l,\beta }}\left[ {\left| {{{\cal D}_1}} \right|} \right]}} \\
&= {e^{ - {\lambda _o}\frac{2}{\pi }{\mathbb E}\left[ l \right]d}}.
\end{aligned} \label{PL}
\end{equation}

For the reconfigured LoS link, the probability of IRS availability must be analyzed first. We consider the available IRSs to be the thinning of $\Phi_i$, which is denoted as $\Psi _i$. The probability of thinning is deduced as
\begin{equation}
\setlength\abovedisplayskip{7pt}
\setlength\belowdisplayskip{7pt}
\begin{aligned}
&\!\!\!{p_i}\left( {r,\theta \left| d \right.} \right) \\
&\!\!\!\!\!\!= {\mathbb P}\left[ {{{\cal N}_{{\Phi _o}}}\left| {{\cal D}' \cup {{\cal R}_1}} \right| = 0\left| {{{\cal N}_{{\Phi _o}}}\left| {{{\cal D}_1}} \right| \ne 0} \right.} \right] \times \varepsilon \\
&\!\!\!\!\!\! = {\mathbb P}\left[ {{{\cal N}_{{\Phi _o}}}\left| {\left( {{\cal D}' \cap {{\cal D}_1}} \right) \cup \left( {{{\cal R}_1} \cap {{\cal D}_1}} \right)} \right| = 0\left| {{{\cal N}_{{\Phi _o}}}\left| {{{\cal D}_1}} \right| \ne 0} \right.} \right]\\
 &\!\!\!\!\!\! \times {\mathbb P}\left[ {{{\cal N}_{{\Phi _o}}}\left| {\left( {{\cal D}' \cup {\cal R}_1} \right)\backslash \left[ {\left( {{\cal D}' \cap {\cal D}_1} \right) \cup \left( {{\cal R}_1 \cap {\cal D}_1} \right)} \right]} \right| \!=\! 0} \right] \!\times\! \varepsilon,
\end{aligned} \label{pi0}
\end{equation}
where ${\cal N}_{{{\Phi _o}}} \left| \cdot \right|$ is the number of points of ${\Phi _o}$ in the given region.
Denote the areas of the overlapping regions ${{\cal D}' \cap {{\cal R}_1}}$, ${{\cal D}' \cap {{\cal D}_1}}$ and ${{{\cal D}_1} \cap {{\cal R}_1}}$ as ${\cal S}^{ov}_1$, ${\cal S}^{ov}_2$ and ${\cal S}^{ov}_3$, respectively.
The mean values of the areas of the overlapping regions are derived as
\begin{equation}
\setlength\abovedisplayskip{7pt}
\setlength\belowdisplayskip{7pt}
\begin{aligned}
{{\mathbb E}_{l,{\beta}}}\left[ {{\cal S}_c^{ov}} \right] &= {{\mathbb E}_{l,{\beta}}}\left[ {\frac{1}{2}{l^2}\frac{{\sin \left( {{\omega _c} - {\theta _c}} \right)\sin \left( {{\omega _c}} \right)}}{{\sin \left( {{\theta _c}} \right)}}} \right]\\
 &\mathop   = \limits^{\left( a \right)}     \frac{1}{{2\pi }}{\mathbb E}\left[ {{l^2}} \right]\!\left( {\frac{1}{2} \!+\! \frac{{\pi  - {\theta _c}}}{2}\cot {\theta _c}} \right)\!,c \!\in\! \left\{ {1,2,3} \right\},
\end{aligned}
\end{equation}
where $\theta_c, c \in \left\{ {1,2,3} \right\}$ are angles between two of connecting lines formed by BS, IRS, and user, ${\omega _c}, c \in \left\{ {1,2,3} \right\}$ are angles of one of the connecting lines formed by BS, IRS and user relative to the blockage as shown in Fig. 10, (a) holds because the difference between ${\omega _c}$ and $\beta$ remains constant. $\theta_3$ is equivalent to $\theta$ due to the definition of $\theta$.

\begin{figure}[t]
\centerline{\includegraphics[width=1\linewidth]{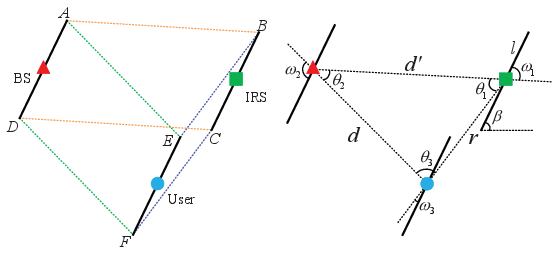}}
\vspace{-0.4cm}
\caption{Link blocking analysis for the proof of {\it Lemma 1}. The quadrilaterals AEFD, BCFE and ABCD are denoted as ${\cal D}_1$, ${\cal R}_1$ and ${\cal D}'$, respectively.}
\vspace{-0.6cm}
\end{figure}

Subsequently, the first term in Eq. \eqref{pi0} is the probability that there is no midpoint of blockage in the regions where ${\cal R}_1$ overlaps with ${\cal D}_1$ and ${\cal D}'$ overlaps with ${\cal D}_1$, provided that there is at least a blockage in ${\cal D}_1$, which is derived as
\begin{equation}
\setlength\abovedisplayskip{1pt}
\setlength\belowdisplayskip{1pt}
\begin{aligned}
&{\mathbb P}\left[ {{{\cal N}_{{\Phi _o}}}\left| {\left( {{\cal D}' \cap {{\cal D}_1}} \right) \cup \left( {{{\cal R}_1} \cap {{\cal D}_1}} \right)} \right| = 0\left| {{{\cal N}_{{\Phi _o}}}\left| {{{\cal D}_1}} \right| \ne 0} \right.} \right]\\
& \mathop  = \limits^{\left( a \right)} \frac{{{\mathbb P}\left[ {{{\cal N}_{{\Phi _o}}}\left| {\left( {{\cal D}' \cap {{\cal D}_1}} \right) \cup \left( {{{\cal R}_1} \cap {{\cal D}_1}} \right)} \right| = 0,{{\cal N}_{{\Phi _o}}}\left| {{{\cal D}_1}} \right| \ne 0} \right]}}{{{\mathbb P}\left[ {{{\cal N}_{{\Phi _o}}}\left| {{{\cal D}_1}} \right| \ne 0} \right]}}\\
& \mathop  = \limits^{\left( b \right)}  \frac{{{e^{ \!- {\lambda _o}{{\mathbb E}_{l,\beta }}\left[\! { {{\cal S}^{ov}_2} + {{\cal S}^{ov}_3}} \!\right]}}\!\! \left( {1 \!-\! {e^{ - {\lambda _o}{{\mathbb E}_{l,\beta }}\left[\! {\left|\! {{{\cal D}_1}} \!\right| - {{\cal S}^{ov}_2} - {{\cal S}^{ov}_3}} \!\right]}}} \right)}}{{1 \!-\! {e^{ - {\lambda _o}{{\mathbb E}_{l,\beta }}\left[ {\left| {{{\cal D}_1}} \right|} \right]}}}}\\
& = \frac{{{e^{ - {}\frac{\lambda _o}{{2\pi }}{\mathbb E}\left[ {{l^2}} \right]\left[ {\sum\limits_{c \in \left\{ {2,3} \right\}}^{} {\left( {\frac{1}{2} + \frac{{\pi  - {\theta _c}}}{2}\cot {\theta _c}} \right)} } \right]}} - {e^{ - \frac{2 {\lambda _o}}{\pi }{\mathbb E}\left[ l \right]d}}}}{{1 - {e^{ - \frac{2{\lambda _o}}{\pi }{\mathbb E}\left[ l \right]d}}}},
\end{aligned}\label{pi1}
\end{equation}
where (a) follows the multiplication law of the conditional probability and (b) approximates the area of the overlapping region of ${\cal D}'$, ${\cal D}_1$, and ${\cal R}_1$ as 0. 
The second term in  Eq. \eqref{pi0} is the probability of no midpoint of blockage in the region of ${\cal R}_1$ and ${\cal D}'$ not coinciding with ${\cal D}_1$, which is derived as
\begin{equation}
\setlength\abovedisplayskip{7pt}
\setlength\belowdisplayskip{7pt}
\begin{aligned}
&{\mathbb P}\left[ {{{\cal N}_{{\Phi _o}}}\left| {\left( {{\cal D}' \cup {{\cal R}_1}} \right)\backslash \left [ \left( {{\cal D}' \cap {{\cal D}_1}} \right) \cup \left( {{{\cal R}_1} \cap {{\cal D}_1}} \right) \right ]}  \right| = 0} \right]\\
& \mathop  = \limits^{\left( a \right)}  {e^{ - {\lambda _o}{{\mathbb E}_{l,\beta }}\left[ {\left| {{\cal D}'} \right| + \left| {{{\cal R}_1}} \right| - {{\cal S}^{ov}_1}-{{\cal S}^{ov}_2}-{{\cal S}^{ov}_3}} \right]}}\\
& = {e^{ - {\lambda _o}\left( {\frac{2}{\pi }{\mathbb E}\left[ l \right]\left( {d' + r} \right) - \frac{1}{{2\pi }}{\mathbb E}\left[ {{l^2}} \right]\left[ {\sum\limits_{c = 1}^3 {\left( {\frac{1}{2} + \frac{{\pi  - {\theta _c}}}{2}\cot {\theta _c}} \right)} } \right]} \right)}},
\end{aligned} \label{pi2}
\end{equation}
where (a) approximates the area of the overlapping region of ${\cal D}'$, ${\cal D}_1$, and ${\cal R}_1$ as zero. 
The third term in Eq. \eqref{pi0} is the probability that the user and the BS are on the same side of the IRS (one of the conditions for the IRS to be able to build a reconfigured LoS link), which is given by
\begin{equation}
\setlength\abovedisplayskip{3pt}
\setlength\belowdisplayskip{3pt}
\varepsilon  = 1 - \frac{{{\theta _1}}}{\pi }.\label{pi3}
\end{equation}
Substituting \eqref{pi1}, \eqref{pi2}, and \eqref{pi3} into \eqref{pi0}, we obtain the expressions for ${p_i}\left( {r,\theta \left| d \right.} \right)$.

To obtain the closed-form expression for densities, we further obtain an approximation of ${p_i}\left( {r,\theta \left| d \right.} \right)$. Based on \eqref{pi0}, we have
\begin{equation}
\setlength\abovedisplayskip{3pt}
\setlength\belowdisplayskip{3pt}
\begin{aligned}
& {p_i}\left( {r,\theta \left| d \right.} \right) \\ & \mathop  \approx \limits^{\left( a \right)} \left\{ \begin{array}{ll}
{\mathbb P}\left[ {{{\cal N}_{{\Phi _o}}}\left| {{{\cal D}^\prime } \cup {{\cal R}_1}} \right| = 0\left| {{{\cal N}_{{\Phi _o}}}\left| {{{\cal D}_1}} \right| \ne 0} \right.} \right] \times \varepsilon ,&r \le {\mathbb E}\left[ l \right]\\
{\mathbb P}\left[ {{{\cal N}_{{\Phi _o}}}\left| {{{\cal D}^\prime } \cup {{\cal R}_1}} \right| = 0} \right] \times \varepsilon ,&r > {\mathbb E}\left[ l \right]
\end{array} \right. \\
 & 
  \mathop  \approx \limits^{\left( b \right)} \left\{ \begin{array}{ll}
0,&r \le {\mathbb E}\left[ l \right]\\
\frac{1}{2}{e^{ - \frac{{2{\lambda _o}}}{\pi }{\mathbb E}\left[ l \right]\left( {d + r} \right)}},&r > {\mathbb E}\left[ l \right]
\end{array} \right.,
\end{aligned}
\end{equation}
where (a) is from the fact that event ${{{\cal N}_{{\Phi _o}}}\left| {{{\cal D}^\prime } \cup {{\cal R}_1}} \right| = 0}$ and  ${{{\cal N}_{{\Phi _o}}}\left| {{{\cal D}_1}} \right| \ne 0}$ are almost independent when $r > {\mathbb E}\left[ l \right]$ and $\varepsilon $ takes 0.5 due to symmetry, (b) is obtained by $d' \approx d$ [3], $\left| {{{\cal D}^\prime } \cap {{\cal R}_1}} \right| \ll \left| {{{\cal D}^\prime }} \right| + \left| {{{\cal R}_1}} \right|$ when $r > {\mathbb E}\left[ l \right]$, and ${\mathbb P}\left[ {{{\cal N}_{{\Phi _o}}}\left| {{{\cal D}^\prime } \cup {{\cal R}_1}} \right| = 0} \right] \approx 0$ if  $r  \le  {\mathbb E}\left[ l \right]$ and ${{{\cal N}_{{\Phi _o}}}\left| {{{\cal D}_1}} \right| \ne 0}$.

Therefore, when the propagation path between the user and BS at distance $d$ is blocked, the probability that an IRS exists to establish a reconfigured LoS link is given by
\begin{equation}
\setlength\abovedisplayskip{7pt}
\setlength\belowdisplayskip{3pt}
\begin{aligned}
{\mathbb P}\left[ {{\Psi _i} \ne \emptyset \left| d \right.} \right] &= 1 - {e^{ - {\lambda _i}\int\limits_{ - \pi }^\pi \! {\int\limits_0^D {{p_i}\left( {r,\theta \left| d \right.} \right)r{\rm{d}}r{\rm{d}}\theta } } }}\\
&   \approx  1 - {e^{ - {\lambda _i}{{\widehat p}_i}\left( {0,D,d} \right)}},
\label{PIRS}
\end{aligned}
\end{equation}
where ${{{\widehat p}_i}\left( {0,D,d} \right)}$ is the integral result with the approximation of ${p_i}\left( {r,\theta \left| d \right.} \right)$ brought in.

Then, the probability of the reconfigured LoS link between the user and the BS at distance $d$ is expressed as
\begin{equation}
\setlength\abovedisplayskip{4pt}
\setlength\belowdisplayskip{4pt}
{p_R}\left( d \right) = \left[ {1 - {p_L}\left( d \right)} \right]{\mathbb P}\left[ {{\Psi _i} \ne \emptyset \left| d \right.} \right]. \label{PR}
\end{equation}
For the NLoS link, the propagation path between the user and BS is blocked, and there is no available IRS; thus, the probability of the NLoS link between the user and BS at distance $d$ is given by
\begin{equation}
\setlength\abovedisplayskip{4pt}
\setlength\belowdisplayskip{4pt}
{p_N}\left( d \right) = \left[ {1 - {p_L}\left( d \right)} \right]\left( {1 - {\mathbb P}\left[ {{\Psi _i} \ne \emptyset \left| d \right.} \right]} \right). \label{PN}
\end{equation}
Based on the marked point process, the densities of BSs at LoS, NLoS, and reconfigured LoS for user are given by
\begin{equation}
\setlength\abovedisplayskip{4pt}
\setlength\belowdisplayskip{4pt}
{\lambda _{b,k}}\left( d \right) = {\lambda _b}{p_k}\left( d \right),\ \ k \in \left\{ {L,N,R} \right\}. \label{lambk}
\end{equation}
By substituting \eqref{PL} and \eqref{PIRS} into \eqref{PR} and \eqref{PN}, we obtain the expressions of ${p_R}\left( d \right)$ and ${p_N}\left( d \right)$ respectively. From ${p_k}\left( d \right),k \in \left\{ {L,K,R} \right\}$ and \eqref{lambk}, we have {\it Lemma 1}.

Note that although only one specific case is shown in Fig. 10, our derivations consider all cases in which the variables are within their range of values.

\subsection{Proof of Lemma 2}

Using the probability ${p_i}\left( {r,\theta \left| d \right.} \right)$ that the IRS at a distance of $r$ and an angle of $\theta$ relative to the user can build a reconfigured LoS, the cdf of the distance between the user and its serving IRS is derived by
\begin{equation} \label{Fr01}
\setlength\abovedisplayskip{3pt}
\setlength\belowdisplayskip{3pt}
\begin{aligned}
{\cal F}{_{{r_1}}}\left( {{r_1}\left| d \right.} \right) &= {\mathbb P}\left[ {{{\cal N}_{{\Psi _i}}}\left| {\cal {C}}\left( {{r_1}} \right)  \right| \ne 0\left| {{\Psi _i} \ne \emptyset ,d} \right.} \right]\\
 &= \frac{{{\mathbb P}\left[ {{{\cal N}_{{\Psi _i}}}\left| {\cal {C}}\left( {{r_1}} \right)  \right| \ne 0\left| d \right.} \right]}}{{{\mathbb P}\left[ {{\Psi _i} \ne \emptyset \left| d \right.} \right]}},
\end{aligned}
\end{equation}
where ${\cal {C}}\left( {{r_1}} \right) $ is a circular region centered on the user's location with $r_1$ as the radius and $\Psi_i$ is the set of available IRSs for the typical user. Then,
\begin{equation} \label{Fr02}
\setlength\abovedisplayskip{1pt}
\setlength\belowdisplayskip{1pt}
\begin{aligned}
{\mathbb P}\left[ {{{\cal N}_{{\Psi _i}}}\left| {\cal {C}}\left( {{r_1}} \right) \right| \ne 0\left| d \right.} \right] &= 1 - {e^{ - {\lambda _i}\int\limits_{ - \pi }^\pi  {\int\limits_0^{{r_1}} {{p_i}\left( {r,\theta \left| d \right.} \right)r{\rm{d}}r{\rm{d}}\theta } } }}\\
 & \approx 1 - {e^{^{ - {\lambda _i}{{\widehat p}_i}\left( {0,{r_1},d} \right)}}},
\end{aligned}
\end{equation}
where ${{{\widehat p}_i}\left( {0,r_1,d} \right)}$ is the integral result of the approximation of ${p_i}\left( {r,\theta \left| d \right.} \right)$ obtained in {\it Lemma 1}.

The expression for ${{p_i}\left( {r,\theta \left| d \right.} \right)}$ and ${{\mathbb P}\left[ {{\Psi _i} \ne \emptyset \left| d \right.} \right]}$ are derived in the proof of {\it Lemma 1}. Substituting \eqref{PIRS} and \eqref{Fr02} into \eqref{Fr01}, we obtain \eqref{Fr0}. 

The pdf of the distance between the user and its serving IRS is then derived as
${f_{{r_1}}}\left( {{r_1}\left| d \right.} \right) = {{{\rm{d}}{{\cal F}_{{r_1}}}\left( {{r_1}\left| d \right.} \right)} \mathord{\left/
 {\vphantom {{{\rm{d}}{{\cal F}_{{r_1}}}\left( {{r_1}\left| d \right.} \right)} {d{r_1}}}} \right.
 \kern-\nulldelimiterspace} {{\rm{d}}{r_1}}} $.

\subsection{Proof of Theorem 1}

The user's serving link at the beginning of the unit of time is in LoS, NLoS, or reconfigured LoS, implying that the BS in ${\Phi _{b,L}}$, ${\Phi _{b,N}}$, or ${\Phi _{b,R}}$, which is closest to the user, provides the maximum received power strength compared with all other BSs.

For the BS in ${\Phi _{b,L}}$, ${\Phi _{b,N}}$, or ${\Phi _{b,R}}$, which is closest to the user, the distance from the typical user is denoted as $d^k , k \in \left\{ {L,N,R} \right\}$, and the pdf of $d^k$ is obtained via {\it Lemma 1}, which is deduced as
\begin{equation}
\setlength\abovedisplayskip{2pt}
\setlength\belowdisplayskip{2pt}
\begin{aligned}
{f_{{d^k}}}\left( {{d^k}} \right) &= \frac{{{\rm{d}}{\mathbb P}\left[ {{{\cal N}_{{\Phi _{b,k}}}}\left| {\cal {C}}\left( {{d^k}} \right)  \right| \ne 0} \right]}}{{{\rm{d}}{d^k}}},
\end{aligned}
\end{equation}
where ${\cal {C}}\left( {{d^k}} \right) $ is a circular region centered on the user's location with $d^k$ as its radius.

Considering the closest BS in ${{\Phi _{b,L}}}$ is $n$ and the distance is $d^L$, the probability that a typical user is associated with BS $n$ is given by
\begin{equation}
\setlength\abovedisplayskip{3pt}
\setlength\belowdisplayskip{2pt}
\begin{aligned}
&{{\cal A}_{L,n}}
 = {\mathbb P}\left[ {\left. {{P _{L,n}} > \mathop {\max }\limits_{m \in {\Phi _{b,k}},k \in \left\{ {N,R} \right\}} {P _{k,m}}} \right|{d^L}} \right]\\
 &\!\!\mathop  = \limits^{\left( a \right)} {\mathbb P}\left[ {\left. {{P _L}\left( {{d^L}} \right) \!\!>\!\! {P _N}\left( {{d^N}} \right)} \right|{d^L}} \right] \!\times\! {\mathbb P}\left[ {\left. {{P _L}\left( {{d^L}} \right) \!\!>\!\! {P _R}\left( {{d^R},{r_1}} \right)} \right|{d^L}} \right]\\
 &\!\! = {\mathbb P}\left[ \!{\left. {{d^N} \!\!>\!\! {{\left( {\frac{{{K_N}}}{{{K_L}}}} \right)}^{\frac{1}{{{\alpha _N}}}}}{{\left( {{d^L}} \right)}^{\frac{{{\alpha _L}}}{{{\alpha _N}}}}}} \right|\!{d^L}} \right] \!\times\! {\mathbb P}\left[ \!{\left. {{r_1} \!\!>\!\! {{\left( {{K_L}{G_{bf}}} \right)}^{\frac{1}{{{\alpha _L}}}}}\frac{{{d^L}}}{{{d^R}}}} \right|\!{d^L}} \right]\!,
\end{aligned}
\end{equation}
where the IRS-BS distance is approximated as the user-BS distance in (a), as in \cite{b25, b3}.
The first term of the equation is the probability that there is no BS in ${{\Phi _{b,L}}}$ in the circular region centered on the user's location and with radius $ {\left( {\frac{{{K_N}}}{{{K_L}}}} \right)^{\frac{1}{{{\alpha _N}}}}}{\left( {{d^L}} \right)^{\frac{{{\alpha _L}}}{{{\alpha _N}}}}}$ (abbreviated as ${{\widetilde d}^N}$).
The second term of the equation is the probability that all IRSs that establish reconfigurable LoS links with the BS in ${{\Phi _{b,R}}}$ are at a distance greater than ${\left( {{K_L}{G_{bf}}} \right)^{\frac{1}{{{\alpha _L}}}}}\frac{d^L}{{{d^R}}}$ (abbreviated as ${{\widetilde r}^L}$) from the user. Using the null probability of the HPPP \cite{b204}, the probabilities are deduced as
\begin{equation} \label{PP}
\setlength\abovedisplayskip{2pt}
\setlength\belowdisplayskip{2pt}
\begin{aligned}
&{\mathbb P}\left[ {\left. {{d^N} > {\widetilde d^N}} \right|{d^L}} \right] \!=\! \exp \left\{ { - 2\pi \int_0^{{{\widetilde d}^N}} \!\!\!{{\lambda _{b,N}}\left( {{d^N}} \right){d^N}{\rm{d}}{d^N}} } \right\},\\
&{\mathbb P}\left[{\left. {{r_1} > {{\widetilde r}^L}}\right|{d^L}} \right] \!=\! \exp \left\{ { - 2\pi \!\! \int_0^\infty \!\!\! \!{{\lambda _{b,R}}\left( {{d^R}} \right)\!{{\cal F}_{{r_1}}}\!\left( {{{\widetilde r}^L}\left| {{d^R}} \right.} \right){d^R}{\rm{d}}{d^R}} } \right\}.
\end{aligned}
\end{equation}
Substituting \eqref{PP} into ${{\mathbb E}_{{d^L}}}\left[ {{{\cal A}_{L,n}}} \right]$, we obtain the expression for ${{\cal A}_{L}}$. For ${{\cal A}_{N}}$, the derivation steps are similar to those for ${{\cal A}_{L}}$.

For a given $d^R$, the probability that a typical user is associated with a BS $n \in \Phi_{b,R}$ is given by
\begin{equation}
\setlength\abovedisplayskip{2pt}
\setlength\belowdisplayskip{2pt}
\begin{small}
\begin{aligned}
\!\!{{\cal A}_{R,n}} \!&=\!{\mathbb P}\left[ {{P _{R,n}} > \mathop {\max }\limits_{m \in {\Phi _{b,k}},k \in \left\{ {L,N} \right\}} {P _{k,m}}} \right]\\
 &\mathop  = \limits^{\left( a \right)} \!{\mathbb P}\!\left[ {{P _R}\left( {{d^R}\!,{r_1}} \right) \!>\! {P _L}\left( {{d^L}} \right)} \right] \!\times\! {\mathbb P}\left[ {{P _R}\left( {{d^R}\!,{r_1}} \right) \!>\! {P _N}\left( {{d^N}} \right)} \right]\\
 &= \!{\mathbb P}\!\left[ { {r_1} \!<\! {{\left( {{K_L}{G_{bf}}} \right)}\!^{\frac{1}{{{\alpha _L}}}}}\frac{{{d^L}}}{{{d^R}}}} \right] \!\!\times\! {\mathbb P}\!\left[  {{r_1} \!<\! {{\left( {\frac{{K_L^2{G_{bf}}}}{{{K_N}}}} \right)}^{\frac{1}{{{\alpha _L}}}}}\!\!\frac{{{{\left( {{d^N}} \right)}\!^{\frac{{{\alpha _N}}}{{{\alpha _L}}}}}}}{{{d^R}}}} \right]\!,\\[0.5mm]
\end{aligned}
\end{small}
\end{equation}
where the IRS-BS distance is approximated as the user-BS distance in (a), as in \cite{b25, b3}.
The two terms of the equation are the probabilities of existing an IRS that establish reconfigurable LoS links with the BS at $d^R$ are at a distance smaller than ${\left( {{K_L}{G_{bf}}} \right)^{\frac{1}{{{\alpha _L}}}}}\frac{d^L}{{{d^R}}}$ and ${{\left( {\frac{{K_L^2{G_{bf}}}}{{{K_N}}}} \right)}^{\frac{1}{{{\alpha _L}}}}}\!\!\frac{{{{\left( {{d^N}} \right)}\!^{\frac{{{\alpha _N}}}{{{\alpha _L}}}}}}}{{{d^R}}}$ (abbreviated as ${{\widetilde r}^L}$ and ${{\widetilde r}^N}$) from the user for given $d^L$ and $d^N$. Using the null probability of the HPPP \cite{b204}, the probabilities are deduced as
\begin{equation} \label{PPr}
\setlength\abovedisplayskip{7pt}
\setlength\belowdisplayskip{7pt}
\begin{aligned}
&{\mathbb P}\!\left[{\left. \!{{r_1} \!\!<\! {{\widetilde r}^L}}\right|\!{d^R}} \right]\!\!=\! \exp \left\{ { - 2\pi \!\!\int_0^\infty  \!\!\!\!\!{{\lambda _{b,L}}\left( {{d^L}}\! \right)\!\!\left[ {1 \!- {{\cal F}_{{r_1}}}\left( {{{\widetilde r}^L}\!\left| {{d^R}} \right.} \right)} \right]\!{d^L}{\rm{d}}{d^L}} } \!\right\}\!,\\
&{\mathbb P}\!\left[{\left. \!{{r_1} \!\!<\! {{\widetilde r}^N}}\right|\!{d^R}} \right]\!\!=\! \exp \left\{ { - 2\pi \!\!\!\int_0^\infty  \!\!\!\!\!{{\lambda _{b,N}}\left( {{d^N}} \!\right)\!\!\left[ \!{1 \!-\!\! {{\cal F}_{{r_1}}}\!\!\left( {{{\widetilde r}^N}\!\left| {{d^R}} \right.} \right)} \right]\!{d^N}{\rm{d}}{d^N}} }\! \right\}\!.
\end{aligned}
\end{equation}
Substituting \eqref{PPr} into ${{\mathbb E}_{{d^R}}}\left[ {{{\cal A}_{R,n}}} \right]$, we obtain the expression for ${{\cal A}_{R}}$.

\subsection{Proof of Theorem 2}

The cdf of ${x_1^k}$ is equal to the probability that ${d^k}<{x_1^k}$ provided that the serving link is in LoS, NLoS, or reconfigured  LoS. Taking $k = L$ as an example, the cdf of ${x_1^L}$ is derived as
\begin{equation}
\setlength\abovedisplayskip{7pt}
\setlength\belowdisplayskip{7pt}
\begin{aligned}
&{{\cal F}_{x_1^k}}\left( {x_1^k} \right) = {\mathbb P}\left[ {\left.{d^k} < x_1^k \right| k=L } \right] = \frac{{{\mathbb P}\left[ {{d^k} < x_1^k,k=L} \right]}}{{{{\cal A}_L}}}\\
 &= \frac{{\int_0^{x_1^L} \! {{\mathbb P}\!\left[ {{P _L}\left( {{d^L}} \right) \!>\! \max \left\{ {{P _N}\left( {{d^N}} \right)\!,{P _R}\left( {{d^R},{r_1}} \right)} \right\}} \right]\!{f_{{d^L}}}\!\!\left( {{d^L}} \right)\!{\rm{d}}{d^L}} }}{{{{\cal A}_L}}}\\
 &= \frac{1}{{{{\cal A}_L}}}\int_0^{x_1^L}  {{\mathbb P}\left[ \left.{{d^N} > {{\widetilde d}^N}} \right| d^L \right]{\mathbb P}\left[ \left. {{r_1} > {{\widetilde r}^N}} \right| d^L \right]{f_{{d^L}}}\left( {{d^L}} \right){\rm{d}}{d^L}} .
\end{aligned}
\end{equation}
Further derivations of ${\mathbb P}\left[ \left.{{d^N} > {{\widetilde d}^N}} \right| d^L \right]$ and ${\mathbb P}\left[ \left. {{r_1} > {{\widetilde r}^N}} \right| d^L \right]$ can be found in the derivation of {\it Theorem 1}. ${{\cal F}_{x_1^k}}\left( {x_1^k} \right)$  for $k \in \left\{ N, R\right\}$ can be derived using derivation steps similar to those for ${{\cal F}_{x_1^L}}\left( {x_1^L} \right)$. Then, the pdf of $x_1^k$ is derived as
${f_{x_1^k}}\left( {x_1^k} \right) = {{{\rm{d}}{{\cal F}_{x_1^k}}\left( {x_1^k} \right)} \mathord{\left/
 {\vphantom {{{\rm{d}}{ {\cal F}_{x_1^k}}\left( {x_1^k} \right)} {{\rm{d}}x_1^k}}} \right.
 \kern-\nulldelimiterspace} {{\rm{d}}x_1^k}}$.

\begin{figure}[t]
\centerline{\includegraphics[width=0.65\linewidth]{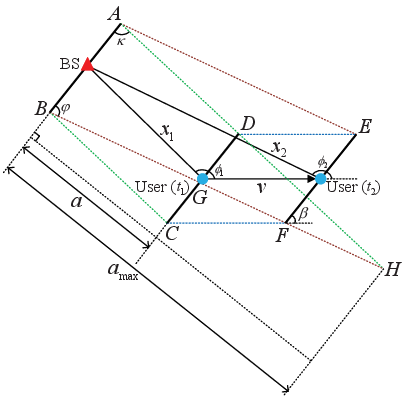}}
\vspace{-0.2cm}
\caption{Link blocking analysis for the proof of {\it Lemma 3}. The blue dots represent the locations of the typical user before and after the movement, and the red triangle represents the BS the user is connected to (can also represent IRS). The quadrilaterals ABCD, ABFE, ABGD and DGFE are denoted as ${\cal S}^{t_1}$, ${\cal S}^{t_2}$, ${\cal S}^{t_1,t_2}_\cap$ and ${{{\cal S}_\emptyset }}$, respectively. $a$ and $a_{\max}$ are the heights of trapezoid ABGD and triangle ABH respectively.}
\vspace{-0.5cm}
\end{figure}

\subsection{Proof of Lemma 3}

As shown in Fig. 11, we define a few regions for the derivation. ${{\cal S}^{t_1}}$ and ${{\cal S}^{t_2}}$ are the sets of all possible locations of the blockage midpoint that block the link when the user is at the initial location and after movement. The region in which ${{\cal S}^{t_1}}$ and ${{\cal S}^{t_2}}$ overlap is denoted by ${{\cal S}_ \cap ^{{t_1},{t_2}}}$. For a given ${\phi _1}$, $x_1$, $\beta$, and $l$, the area ${{\cal S}_ \cap ^{{t_1},{t_2}}}$ is given by
\begin{equation}\label{Areas1}
\left|{\cal S}_ \cap ^{{t_1},{t_2}}\right| \!=\! \left\{ \begin{array}{ll}
\!\!0,&{\phi _1} \!\le\! \beta  \!\le\! {\phi _2}\\
\!\!{la\left( {1 - \frac{1}{2}a{{\dot a}_{\max }}} \right)},&{\rm others}
\end{array} \right.\!\!\!.
\end{equation}
where ${{\dot a}_{\max }} = \left\{ \begin{array}{ll}
0,&{\phi _1},{\phi _2} \in \left\{ {0,\pi } \right\},\beta  = \frac{\pi }{2}\\
\frac{1}{{{a_{\max }}}},&{\rm others}
\end{array} \right.$, $a$ and $a_{\max}$ are the heights of triangle ABH and trapezoid ABGD, as shown in Fig. 11.

The union of ${{\cal S}^{t_2}}$ and the set of all possible locations of the blockage midpoint blocking the user's movement trajectory is denoted by ${{{\cal S}_\emptyset }}$. For a given ${\phi _1}$, $x_1$, $\beta$, and $l$, the area ${{{\cal S}_\emptyset }}$ is given by
\begin{equation}
 \!\left| {{\cal S}_\emptyset }\right| \!\!=\!\! \left\{ \begin{array}{ll}\!\!
\frac{1}{2}\!\left( {{l^2}\sin \beta \cos \beta  \!-\! \frac{{{l^2}{{\sin }^2}\beta }}{{\tan {\phi _2}}}} \right)\!,\\  
&\!\!\!\!\!\!\!\!\!\!\!\!\!\!\!\!\!\!\!\!\!\!\!\!\!\!\!\!\!\!\!\!\!\!\!\!\!\!\!\!\!\!\!\!\!\!\!\!\!\!\!\!\!\!
l\!\cos\! \beta  \!-\! \frac{{l\sin \beta }}{{\tan {\phi _2}}} \!\le\!\! v,\beta  \!<\! {\phi _2}, {\phi _2} \notin \left\{ {0,\frac{\pi }{2},\pi } \right\}
\\
lv\sin \beta  \!-\! \frac{{{v^2}\sin \beta }}{{2\left( {\cos \! \beta  \!-\! \frac{{\sin \beta }}{{\tan {\phi _2}}}} \right)}},
\\ 
&\!\!\!\!\!\!\!\!\!\!\!\!\!\!\!\!\!\!\!\!\!\!\!\!\!\!\!\!\!\!\!\!\!\!\!\!\!\!\!\!\!\!\!\!\!\!\!\!\!\!\!\!\!\!
l\!\cos \! \beta  \!-\! \frac{{l\sin \beta }}{{\tan {\phi _2}}} \!>\!\! v,\beta  \!<\! {\phi _2}, {\phi _2} \notin \left\{ {0,\frac{\pi }{2},\pi } \right\}
\\
 lv\sin \beta ,&\!\!\!\!\!{\phi _2} \in \left\{ {0,\pi } \right\}\\
\frac{1}{2}{l^2}\sin \beta \cos \beta ,&\!\!\!\!\!{\phi _2} = \frac{\pi }{2},l\cos \beta  \le v\\
 lv\sin \beta  - \frac{1}{2} {v^2} \tan \beta ,&\!\!\!\!\!{\phi _2} = \frac{\pi }{2},l\cos \beta  > v
\\
0,&\!\!\!\!\! \beta  \ge  {\phi _2}   
\end{array} \right.\!\!\!,\!\!\! \label{Areas2}
\end{equation}
The considerations of ${{{\cal S}_\emptyset }}$ show the effect of the modified random walk model proposed in the analysis.

The probability that a single link initially in the LoS maintains LoS after user movement is deduced as

\begin{equation}\label{PLL}
\setlength\abovedisplayskip{7pt}
\setlength\belowdisplayskip{7pt}
\begin{aligned}
\!\!\!\!\!{\cal P}_{L,L}^{link}\left( {{x_1},{\phi _1}} \right) &= {\mathbb P}\left[ {{{\cal N}_{{\Phi _o}}}\left| {{{\cal S}^{{t_2}}}\backslash {{\cal S}_\emptyset }} \right| = 0\left| {{{\cal N}_{{\Phi _o}}}\left| {{{\cal S}^{{t_1}}}} \right| = 0} \right.} \right]\\
 &= \frac{{{\mathbb P}\left[ {{{\cal N}_{{\Phi _o}}}\left| {{{\cal S}^{{t_2}}}\backslash {{\cal S}_\emptyset }} \right| = 0,{{\cal N}_{{\Phi _o}}}\left| {{{\cal S}^{{t_1}}}} \right| = 0} \right]}}{{{\mathbb P}\left[ {{{\cal N}_{{\Phi _o}}}\left| {{{\cal S}^{{t_1}}}} \right| = 0} \right]}}\\
&= \frac{{{\mathbb P}\left[ {{{\cal N}_{{\Phi _o}}}\left| {{{\cal S}^{{t_1}}} \cup {{\cal S}^{{t_2}}}\backslash {{\cal S}_\emptyset }} \right| = 0} \right]}}{{{\mathbb P}\left[ {{{\cal N}_{{\Phi _o}}}\left| {{{\cal S}^{{t_1}}}} \right| = 0} \right]}}\\
 &\mathop  = \limits^{\left( a \right)} \exp \left\{ { - {\lambda _o}{{\mathbb E}_{l,\beta }}\left[ {\left| {{{\cal S}^{{t_2}}}} \right| - \left| {{\cal S}_ \cap ^{{t_1},{t_2}}} \right| - \left| {{{\cal S}_\emptyset }} \right|} \right]} \right\},
\end{aligned}
\end{equation}
where (a) follows the geometric relations of ${{{\cal S}^{{t_1}}}}$, ${{{\cal S}^{{t_2}}}}$, ${{{\cal S}_\emptyset }}$, and ${{\cal S}_ \cap ^{{t_1},{t_2}}}$. Similarly, we derive ${\cal P}_{N,L}^{link}\left( {{x_1},{\phi _1}} \right)$ as
\begin{equation} \label{PNL}
\setlength\abovedisplayskip{7pt}
\setlength\belowdisplayskip{7pt}
\begin{aligned}
&{\cal P}_{N,L}^{link}\left( {{x_1},{\phi _1}} \right) = {\mathbb P}\left[ {{{\cal N}_{{\Phi _o}}}\left| {{{\cal S}^{{t_2}}}\backslash {{\cal S}_\emptyset }} \right| = 0\left| {{{\cal N}_{{\Phi _o}}}\left| {{{\cal S}^{{t_1}}}} \right| \ne 0} \right.} \right]\\
& = \frac{{{\mathbb P}\left[ {{{\cal N}_{{\Phi _o}}}\left| {{{\cal S}^{{t_2}}}\backslash {{\cal S}_\emptyset }} \right| = 0} \right] \times {\mathbb P}\left[ {{{\cal N}_{{\Phi _o}}}\left| {{{\cal S}^{{t_1}}}\backslash {\cal S}_ \cap ^{{t_1},{t_2}}} \right| \ne 0} \right]}}{{{\mathbb P}\left[ {{{\cal N}_{{\Phi _o}}}\left| {{{\cal S}^{{t_1}}}} \right| \ne 0} \right]}}\\
& = \frac{{\left( {1 \!-\! \exp \left\{ { - {\lambda _o}{{\mathbb E}_{l,\beta }}\!\left[ {\left| {{{\cal S}^{{t_1}}}} \right| \!-\! \left| {{\cal S}_ \cap ^{{t_1},{t_2}}} \right|} \right]} \right\}} \right)} \exp \left\{ { - {\lambda _o}{{\mathbb E}_{l,\beta }}\!\left[ {\left| {{{\cal S}^{{t_2}}}} \right|} \right]} \right\}}{{1 - \exp \left\{ { - {\lambda _o}{{\mathbb E}_{l,\beta }}\!\left[ {\left| {{{\cal S}^{{t_1}}}} \right|} \right]} \right\}}}.
\end{aligned}
\end{equation}
By substituting \eqref{Areas1} and \eqref{Areas2} into \eqref{PLL} and \eqref{PNL}, we have the final expressions of ${\cal P}_{L,L}^{link}\left( {{x_1},{\phi _1}} \right)$ and ${\cal P}_{N,L}^{link}\left( {{x_1},{\phi _1}} \right)$. Obviously, we have ${\cal P}_{L,N}^{link}\left( {{x_1},{\phi _1}} \right)=1-{\cal P}_{L,L}^{link}\left( {{x_1},{\phi _1}} \right)$ and ${\cal P}_{N,N}^{link}\left( {{x_1},{\phi _1}} \right)=1-{\cal P}_{N,L}^{link}\left( {{x_1},{\phi _1}} \right)$.

Note that although only one specific case is shown in Fig. 11, our derivations consider all cases in which the variables are within their range of values.

\subsection{Proof of Theorem 3}

When analyzing the LoS state of the cascaded link between a typical user and the serving BS, the LoS state of the direct user-BS link and the availability of IRSs are considered.
For the case in which the user-BS link is LoS at the beginning of the unit of time, the respective conditions for the link to maintain LoS, transition to NLoS, or reconfigured LoS are
\begin{itemize}
\item The direct user-BS link is still not blocked after movement.

\item The direct user-BS link is blocked after movement and no available IRS exists.

\item The direct user-BS link is blocked after movement but available IRSs exist.
\end{itemize}
In the case of the NLoS link at the beginning of the time unit, the conditions for the link are LoS, NLoS, and reconfigured LoS after movement, which are similar to the case of the LoS link at the beginning.
In the case of the reconfigured LoS link at the beginning, the conditions for the link are LoS, NLoS, and the reconfigured LoS after movement are
\begin{itemize}
\item The direct user-BS link is not blocked after movement.

\item The direct user-BS link remains blocked, original user-IRS link is blocked, and no available IRS exists after movement. 

\item The direct user-BS link remains blocked after movement and available IRSs exist (including cases of original user-IRS link not blocked and original user-IRS link blocked but new available IRS exists).
\end{itemize}
The transition probabilities of LoS states between the user and the serving BS are obtained by determining the product of probabilities based on the corresponding conditions, where the transition probabilities for a single link are given in {\it Lemma 3} and the expression of $p_i \left(r,\theta | d \right)$ is given in {\it Lemma 1}.

\subsection{Proof of Theorem 4}

The conditional probability ${\bar {\cal H} ^{k,j,w}}$ represents the probability of no HOs to a BS with an LoS state of $w$ under the serving link $k$ at the beginning of the unit of time, and the serving link is $j$ after movement for given $x_1^k$ and $\phi_1$. Users do not directly HO to a BS with reconfigured LoS, as no IRS scheduling before user access \cite{b25, b205}; thus we have $w \in \left\{ L, N \right\}$, $k, j \!\in\! \left\{ L, N,R \right\}$.

We define two regions, ${\cal C}^{k,w}_{1,eq}$ and ${\cal C}^{j,w}_{2,eq}$, which represent the impossible and possible locations of the BSs in the LoS state of $w$ that satisfy the HO condition. The impossibility of a BS with an LoS state of $w$ in ${\cal C}^{k,w}_{1,eq}$ is due to the premise that the user is associated with the BS with an LoS state of $k$ at distance $x_1^k$. Thus, we obtain the following equations for the radii $x^{k,w}_{1,eq}$ and $x^{j,w}_{2,eq}$ of ${\cal C}^{k,w}_{1,eq}$ and ${\cal C}^{k,w}_{2,eq}$
\begin{equation}
\setlength\abovedisplayskip{2pt}
\setlength\belowdisplayskip{2pt}
\begin{aligned}
{P _w}\left( {x_{1,eq}^{k,w}} \right) = \left\{ \begin{array}{ll}
{P _k}\left( {x_1^k} \right),&k \in \left\{ {L,N} \right\}\\[0.5mm]
{P _k}\left( {x_1^k,{r_1}} \right),&k = R
\end{array} \right.,\\
{P _w}\left( {x_{2,eq}^{j,w}} \right) = \left\{ \begin{array}{ll}
{P _k}\left( {x_2^j} \right),&k \in \left\{ {L,N} \right\}\\
{P _k}\left( {x_2^j,{r_2}} \right),&k = R
\end{array} \right..
\end{aligned}
\end{equation}
By transforming this equation, we obtain the expressions for $x^{k,w}_{1,eq}$ and $x^{j,w}_{2,eq}$, where the IRS-BS distance is approximated as the user-BS distance, as in \cite{b25, b3}. Subsequently, $\bar {\mathcal H}^{k,j,w}$ is derived as
\begin{equation}
\setlength\abovedisplayskip{2pt}
\setlength\belowdisplayskip{2pt}
\begin{aligned}
&{\bar {\cal H}^{k,j,w}} = {\mathbb P}\left[ {{{\cal N}_{{\Phi _{b,w}}}}\left| {{\cal C}_{2,eq}^{j,w}/{\cal C}_{1,eq}^{k,w}} \right| = 0} \right]\\
& = {{\mathbb E}_{{r_1},{\phi '_1}}}\left[ { \exp \left\{ { - \iint_{{\cal C}_{2,eq}^{j,w}/{\cal C}_{1,eq}^{k,w}} {{\lambda _{b,w}}\left( x \right)x{\rm{d}}\theta {\rm{d}}x} } \right\}} \right],
\end{aligned}
\end{equation}
where the ranges of integration for $d$ and $\theta$ follow the geometrical relations of ${{\cal C}_{2,eq}^{j,w}}$ and ${{\cal C}_{1,eq}^{k,w}}$, and the means of $r_1$ and $\phi'_1$ are omitted when those random variables are not involved.

By combining the user-BS distance distribution of $x^k_1$ in {\it Theorem 2}, the probabilities of transitions after movement ${{\cal P}_{k,j}^{BS}}$ in {\it Theorem 3}, the probabilities of the initial LoS state of the serving link ${{{\cal A}_k}}$ in {\it Theorem 1}, we obtain the average HO probability in IRS-aided networks considering blockage effects.

\end{appendix}

\vspace{12pt}

\end{document}